\documentclass[12pt]{article}

\usepackage{url, enumerate, algorithmic, algorithm, natbib, amsmath, 
amsfonts, amssymb, amsthm, graphicx, dsfont, titling, float}
\usepackage[margin=1in]{geometry}
\usepackage[affil-it]{authblk}
\newtheorem{prop}{Proposition}

\makeatletter 
\def\@maketitle{ 
\newpage 
\null 
\vskip 2em 
\begin{center} 
\let \footnote \thanks 
{\Large \bfseries \@title \par } 
\vskip 1.5em 
{ \normalsize 
\lineskip .5em 
\begin{tabular}[t]{c} 
\@author 
\end{tabular}\par} 
\vskip 1em 
{\normalsize \@date} 
\end{center} 
\par 
\vskip 1.5em} 
\makeatother

\renewcommand{\theequation}{\arabic{section}.\arabic{equation}}

\begin{document}
\title{Are Clusterings of Multiple Data Views Independent?}

\author{Lucy L. Gao \thanks{\texttt{lucygao@uw.edu}}} 
\affil{Department of Biostatistics, University of Washington}

\author{Jacob Bien} 
\affil{Department  of Data Sciences and Operations, University of Southern California}

\author{Daniela Witten} 
\affil{Departments of Statistics and Biostatistics, University of Washington}

\date{\today} 

\maketitle

\begin{abstract}
{In the Pioneer 100 (P100) Wellness Project \citep{price2017wellness}, multiple types of data are collected on a single set of healthy participants at multiple timepoints in order to characterize and  optimize  wellness. One way to do this is to identify clusters, or subgroups, among the participants, and then to tailor  personalized health recommendations to each subgroup. It is tempting to cluster the participants using all of the data types and timepoints, in order to fully exploit the available information. However, clustering the participants based on multiple \emph{data views} implicitly assumes that a single underlying clustering of the participants is shared across all data views. If this assumption does not hold, then clustering the participants using multiple data views may lead to spurious results. In this paper, we seek to evaluate the assumption that there is some underlying relationship among the clusterings from the different data views, by asking the question: are the clusters within each data view \emph{dependent} or \emph{independent}?  We develop a new test for answering this question, which we then apply to clinical, proteomic, and metabolomic data, across two distinct timepoints, from the P100 study. We find that while the subgroups of the participants defined with respect to any single data type seem to be dependent across time, the clustering among the participants based on one data type (e.g. proteomic data) appears not to be associated with the clustering based on another data type (e.g. clinical data).}
{Data integration; Hypothesis testing; Model-based clustering; Multiple-view data.}
\end{abstract}

\section{Introduction}
\label{sec:intro}

Complex biological systems consist of diverse components with dynamics that may vary over time, and so these sytems often cannot be fully characterized by any single type of data, or at any single snapshot in time. Consequently, it has become increasingly common for researchers to collect multiple data sets, or \emph{views}, for a single set of observations. In the machine learning literature, this is known as the \emph{multiple-view} or \emph{multi-view} data setting.

Multiple-view data has been applied extensively to characterize disease, such as in The Cancer Genome Atlas Project \citep{cancer2008comprehensive}.  In contrast, The Pioneer 100 (P100) Wellness Project \citep{price2017wellness} collected multiple-view data from healthy participants to characterize wellness, and to optimize wellness of the participants through personalized healthcare recommendations. One way to do this is to identify subgroups of similar participants using cluster analysis, and then tailor recommendations to each subgroup. 

In recent years, many papers have proposed clustering methods in the multiple-view data setting 
\citep{bickel2004multi, shen2009integrative, kumar2011co,  kirk2012bayesian, lock2013bayesian, gabasova2017clusternomics}. The vast majority of these methods ``borrow strength" across the data views to obtain a more accurate clustering of the observations than would be possible based on a single data view. Implicitly, these methods assume that there is a single  \emph{consensus} clustering shared by all data views.

The P100 data contains many data views; multiple data types (e.g. clinical data and proteomic data) are available at multiple timepoints. Thus, it is tempting to apply consensus clustering methods to identify subgroups of the P100 participants. However, before doing so, it is important to check the assumption that there exists a single consensus clustering. If instead different views reflect unrelated aspects of the participants, then there is no ``strength to be borrowed" across the views,  and it would be better to perform a separate clustering of the observations in each view.  Before attempting cluster analysis of the P100 data, it is critical that we determine which combinations of views have ``strength to be borrowed", and which combinations do not. 

This raises the natural question of how associated the underlying clusterings are in each view.  Suppose we cluster the P100  participants twice, once using their baseline clinical data, and once using their baseline proteomic data.  \emph{Can we tell from the data whether the two views' underlying clusterings are related or unrelated?} Answering this question provides useful information: 
\begin{list}{}{}
\item{\emph{Case 1:}} If the underlying clusterings appear related, then this increases confidence that the clusterings are scientifically meaningful, and offers some support for performing a consensus clustering of the P100 participants that integrates baseline clinical and proteomic views. 
\item{\emph{Case 2:}} If the underlying clusterings appear unrelated, we must consider two explanations. 
\begin{enumerate}
\item Perhaps clinical and proteomic views measure different properties about the participants, and therefore identify complementary (or ``orthogonal'') clusterings. If so, then a consensus clustering is unlikely to provide meaningful results, and may cause us to lose valuable information about the subgroups underlying the individual data views. 
\item Perhaps the subgroups underlying the data views are indeed related, but they appear unrelated due to noise. If so, then we might be skeptical of any results obtained on these very noisy data, whether from consensus clustering or another approach. 
\end{enumerate}
\end{list}
In Case 2, it would not be appropriate to perform consensus clustering. 

To determine from the data whether the two views' clusterings are related or unrelated, it is tempting to apply a clustering procedure (e.g. k-means) to each view, then apply well-studied tests of independence of categorical variables (e.g. the $\chi^2$-test for independence, the $G$-test for independence, or Fisher's exact test) to the estimated cluster assignments. However, such an approach relies on an assumption that the estimated cluster assignments are independent and identically distributed samples from the joint distribution of the cluster membership variables, which is not satisfied in practice.  Thus, there is a need for an approach which takes into account the fact that the clusterings are estimated from the data.

The rest of this paper is organized as follows.  In Section~\ref{sec:model}, we propose a mixture model for two-view data. In Section~\ref{sec:ind}, we use this model to develop a test of the null hypothesis that clusterings on two views of a single set of observations are independent. We explore the performance of our proposed hypothesis test via numerical simulation in Section~\ref{sec:sim}. In Section~\ref{sec:meila}, we connect and compare our proposed hypothesis test to the aforementioned approach of applying the $G$-test for independence to the estimated cluster assignments, and draw connections between this approach and the mutual information statistic \citep{meilua2007comparing}. In Section~\ref{sec:app}, we apply our method to the clinical, proteomic, and metabolomic datasets from the P100 study. In Section~\ref{sec:discuss}, we provide a discussion, which includes the extension to more than two views.  

\section{A Mixture Model for Multiple-View Data}

\label{sec:model}
\subsection{Model Specification}
\label{sec:mixmodel}
In what follows, we consider the case of two data views. We will discuss the extension to more than two views in Section~\ref{sec:discuss}. 

Suppose we have $p_1$ and $p_2$ features in the first and second data view, respectively. For a single observation, let $X^{(1)} \in \mathbb R^{p_1}$ and $X^{(2)} \in \mathbb R^{p_2}$ denote the random vectors corresponding to the two data views and let $Z^{(1)}\in\{1,\ldots,K^{(1)}\}$ and $Z^{(2)}\in\{1,\ldots,K^{(2)}\}$ be unobserved random variables, indicating the latent group memberships of this observation in the two data views.  Here, $K^{(1)}$ and $K^{(2)}$ represent the number of clusters in the two data views, which we assume for now to be known (we will consider the case in which they are unknown in Section \ref{sec:tun}).  We assume that $X^{(1)}$ and $X^{(2)}$ are conditionally independent given the pair of cluster memberships, $(Z^{(1)}, Z^{(2)})$; this assumption is common in the multi-view clustering literature (see e.g. \citealt{bickel2004multi}, \citealt{rogers2008investigating}, \citealt{kumar2011co}, \citealt{lock2013bayesian}, \citealt{gabasova2017clusternomics}). Further, suppose that 
\begin{align}
\label{eq:mixtureMod}
f(X^{(l)} \mid Z^{(l)} = k) = \phi^{(l)} \left(X^{(l)}; \theta^{(l)}_k\right)&\quad~\text{ for
  }\quad 1\leq k \leq K^{(l)}, ~ 1 \leq l \leq 2, \\
 \label{eq:defPi} P(Z^{(1)}=k,Z^{(2)}=k')=\Pi_{kk'} &\quad~\text{ for
  }\quad 1\leq k \leq K^{(1)}, 1 \leq k' \leq K^{(2)},
\end{align}
where $\phi^{(l)}(\cdot; \theta)$ denotes a density function with parameter $\theta$, and $\Pi\in \Delta^{K^{(1)} \times K^{(2)}} \equiv \{ S \in \mathbb R^{K^{(1)} \times K^{(2)}} : S_{kk'} \geq 0, \sum \limits_{k=1}^{K^{(1)}} \sum \limits_{k'=1}^{K^{(2)}} S_{kk'} = 1 \}$.
Equations \eqref{eq:mixtureMod}--\eqref{eq:defPi} are an extension of the finite mixture model \citep{mclachlan2004finite} to the case of two data views. We further assume that each cluster has positive probability,
 i.e. $P(Z^{(1)}=k)>0$ and $P(Z^{(2)}=k')>0$, and so
$\Pi 1_{K^{(2)}} \in \Delta_+^{K^{(1)}}$ and $\Pi^T 1_{K^{(1)}} \in
\Delta_+^{K^{(2)}}$, where $\Delta^K_+ \equiv \{ s \in \mathbb R^{K} : s_k > 0, \sum \limits_{k=1}^K s_k = 1 \}.$ 

Let $\theta^{(1)} \equiv (\theta^{(1)}_1,\ldots,\theta^{(1)}_{K^{(1)}})$ and 
$\theta^{(2)}\equiv (\theta^{(2)}_1,\ldots,\theta^{(2)}_{K^{(2)}})$. The joint density of $X^{(1)}$ and $X^{(2)}$ is 
\begin{align}
  f(X^{(1)},X^{(2)}; \theta^{(1)}, \theta^{(2)}, \Pi)&=\sum_{k,k'}\Pi_{kk'} f(X^{(1)},X^{(2)}  \mid Z^{(1)}=k,Z^{(2)}=k') \nonumber \\
  &= \sum _{k,k'}\Pi_{kk'} f(X^{(1)} \mid Z^{(1)}=k) f(X^{(2)} \mid Z^{(2)}=k') \nonumber \\
&= \sum_{k,k'} \Pi_{kk'} \phi^{(1)}\left(X^{(1)} ; \theta^{(1)}_k \right)
\phi^{(2)}\left(X^{(2)}  ; \theta^{(2)}_{k'}\right), 
\label{eq:jointdensity}
\end{align}
where the second equality follows from conditional independence of
$X^{(1)}$ and $X^{(2)}$ given $Z^{(1)}$ and $Z^{(2)}$, and the last equality
follows from \eqref{eq:mixtureMod}.   

The matrix $\Pi$ governs the
statistical dependence between the two data views.  It will be useful for us to parameterize
$\Pi$ in terms of a triplet $(\pi^{(1)},\pi^{(2)},C)$ that
separates the single-view information from the cross-view information.  
\begin{prop} 
\label{prop:reparam}
 Suppose $\pi^{(1)} \in \Delta^{K^{(1)}}_+$ and $\pi^{(2)} \in \Delta^{K^{(2)}}_+$. Then, 
\begin{gather*} 
\left \{ \Pi \in \Delta^{K^{(1)} \times K^{(2)}} : \Pi 1_{K^{(2)}} = \pi^{(1)}, ~\Pi^T 1_{K^{(1)}} = \pi^{(2)} \right \} 
=  \left \{ \mathrm{diag}(\pi^{(1)}) C \mathrm{diag}(\pi^{(2)}) : C \in \mathcal{C}_{\pi^{(1)}, \pi^{(2)}} \right \},
\end{gather*} 
where $\mathcal{C}_{\pi^{(1)}, \pi^{(2)}} = \{ C \in \mathbb R^{K^{(1)} \times K^{(2)}} : C_{kk'} \geq 0, ~ C \pi^{(2)} = 1_{K^{(1)}}, ~ C^T \pi^{(1)} = 1_{K^{(2)}}\}.$
\end{prop} 
A proof of Proposition \ref{prop:reparam} is given in Appendix \ref{sec:proofreparam} of the Supplementary Materials. 

Proposition \ref{prop:reparam} 
indicates that any matrix $\Pi \in \Delta^{K^{(1)} \times K^{(2)}}$ with $\Pi 1_{K^{(2)}} \in \Delta_+^{K^{(1)}}$ and $\Pi^T 1_{K^{(1)}} \in \Delta_+^{K^{(2)}}$ can be written as the product of  its row sums
$\pi^{(1)}$, its column sums $\pi^{(2)}$, and a matrix $C$.  Therefore, we can rewrite the joint probability density  \eqref{eq:jointdensity} as follows:
\begin{align}
\label{eq:jointdensityC} 
 f(X^{(1)},X^{(2)}; \theta^{(1)}, \theta^{(2)}, \Pi)  &= \sum_{k,k'} \pi^{(1)}_k C_{kk'} \pi^{(2)}_{k'} \phi^{(1)}\left(X^{(1)} ; \theta^{(1)}_k \right) \nonumber 
\phi^{(2)}\left(X^{(2)}  ; \theta^{(2)}_{k'}\right) \\ 
&\equiv f(X^{(1)}, X^{(2)}; \theta^{(1)}, \theta^{(2)}, \pi^{(1)}, \pi^{(2)}, C). 
\end{align} 
In what follows, we will parametrize the density of  $X^{(1)}$ and $X^{(2)}$ in terms of $\theta^{(1)}, \theta^{(2)}, \pi^{(1)}, \pi^{(2)}$, and $C$, rather than in terms of $\theta^{(1)}, \theta^{(2)}$, and $\Pi$. 

The following proposition characterizes the marginal distributions of $X^{(1)}$ and $X^{(2)}$. 
\begin{prop} 
\label{prop:marginals}
Suppose $X^{(1)}$ and $X^{(2)}$ have joint distribution \eqref{eq:jointdensityC}. Then for $l = 1, 2$, $X^{(l)}$ has marginal density given by 
\begin{align} 
\label{eq:marginaldensities} 
f(X^{(l)} ; \theta^{(l)}, \pi^{(l)}) &= \sum_{k=1}^{K^{(l)}} \pi^{(l)}_k \phi^{(l)}_k \left(X^{(l)} ;\theta^{(l)}_k\right).
\end{align} 
\end{prop} 
Proposition \ref{prop:marginals} follows from \eqref{eq:mixtureMod} -- \eqref{eq:defPi}. Proposition \ref{prop:marginals} shows that for $l = 1, 2$, 
$X^{(l)}$ marginally follows a mixture model with parameters $\theta^{(l)}$ and cluster membership probabilities $\pi^{(l)}$. Note that the marginal
density of $X^{(1)}$ does not depend on $\theta^{(2)}, \pi^{(2)}$,
and $C$, and similarly, the marginal density of $X^{(2)}$ does not
depend on $\theta^{(1)}, \pi^{(1)}$, and $C$;   this fact will
be critical to our approach to parameter estimation in Section~\ref{sec:fitmodel}. 

The model described in this section is closely related to several multiple-view mixture models proposed in the literature: see e.g. 
 \citet{rogers2008investigating},  \citet{kirk2012bayesian}, \citet{lock2013bayesian}, and \citet{gabasova2017clusternomics}. However, the focus of those papers is cluster estimation: they do not provide a statistical test of association, and for the most part, impose additional structure on the probability matrix $\Pi$ in order to encourage similarity between the clusters estimated in each data view. By constrast, the focus of this paper is inference: testing for dependence between the clusterings in different data views. The model described in this section is a step towards that goal. 

\subsection{Interpreting $\Pi$}
\label{sec:interpretpi}
\begin{figure}[H]
\hspace{15mm} (i) \hspace{40mm} (ii) \hspace{40mm} (iii) \\
\includegraphics[scale=0.05]{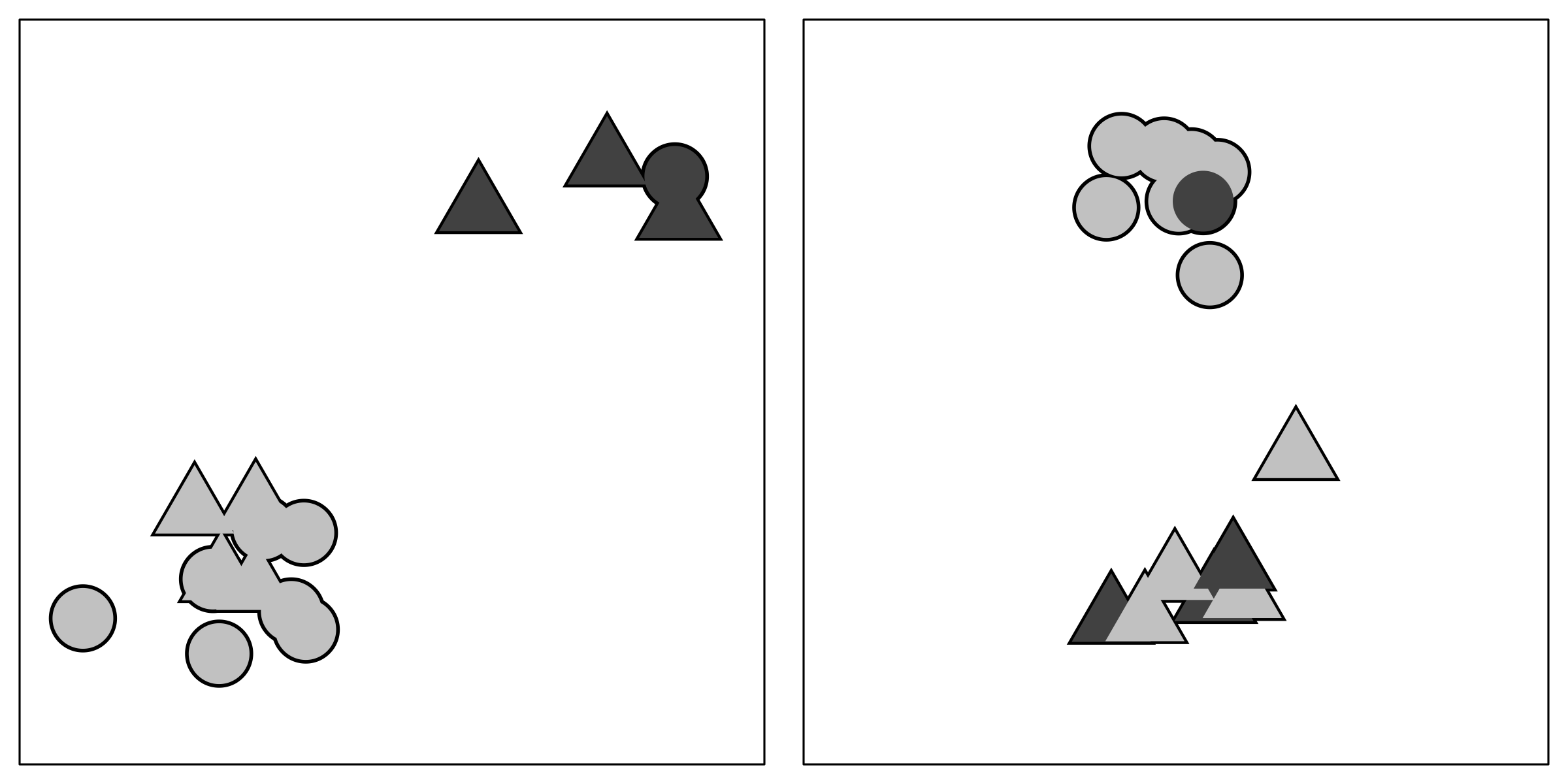} \hspace{5mm}\includegraphics[scale=0.05]{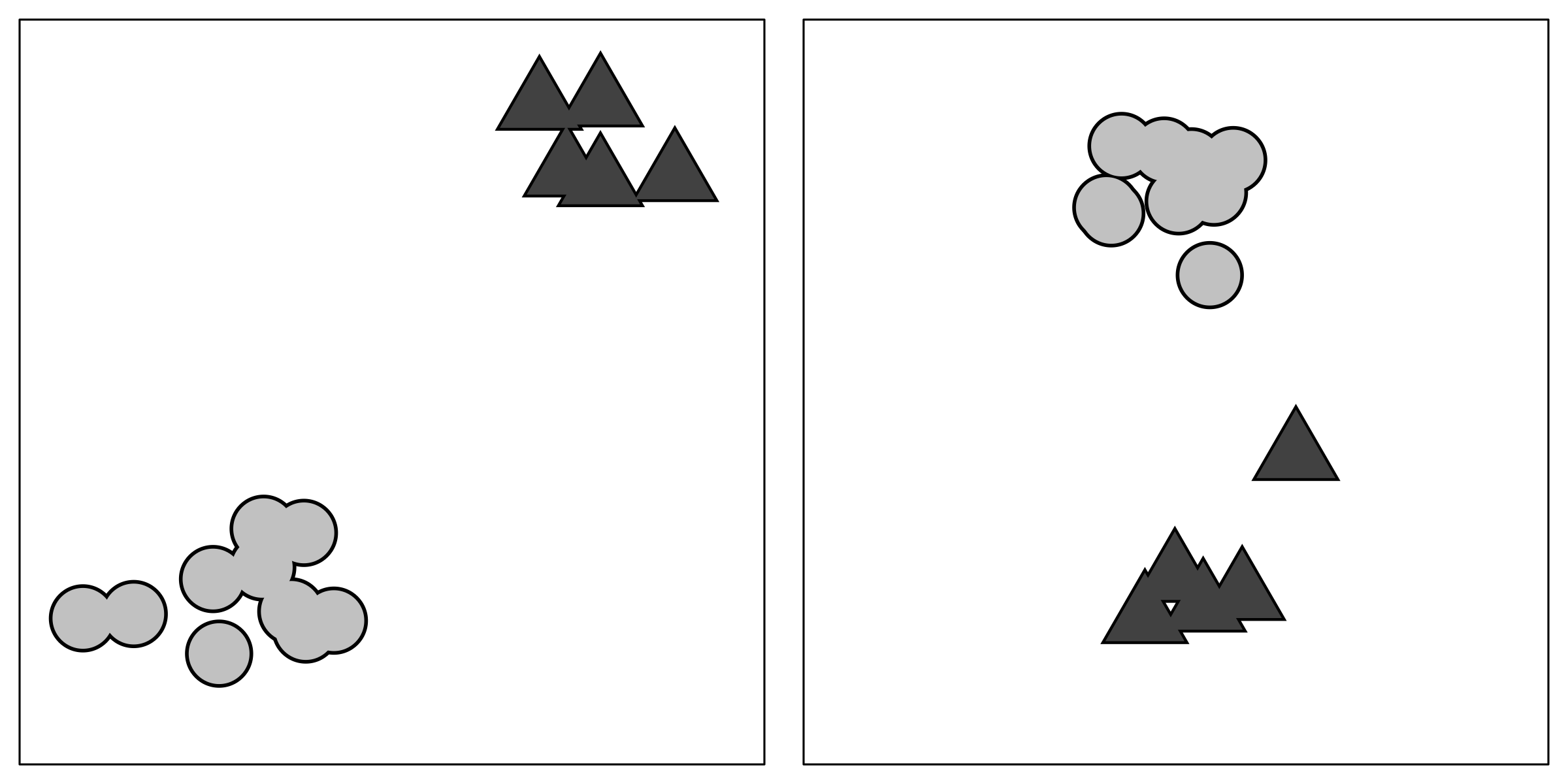} \hspace{5mm}\includegraphics[scale=0.05]{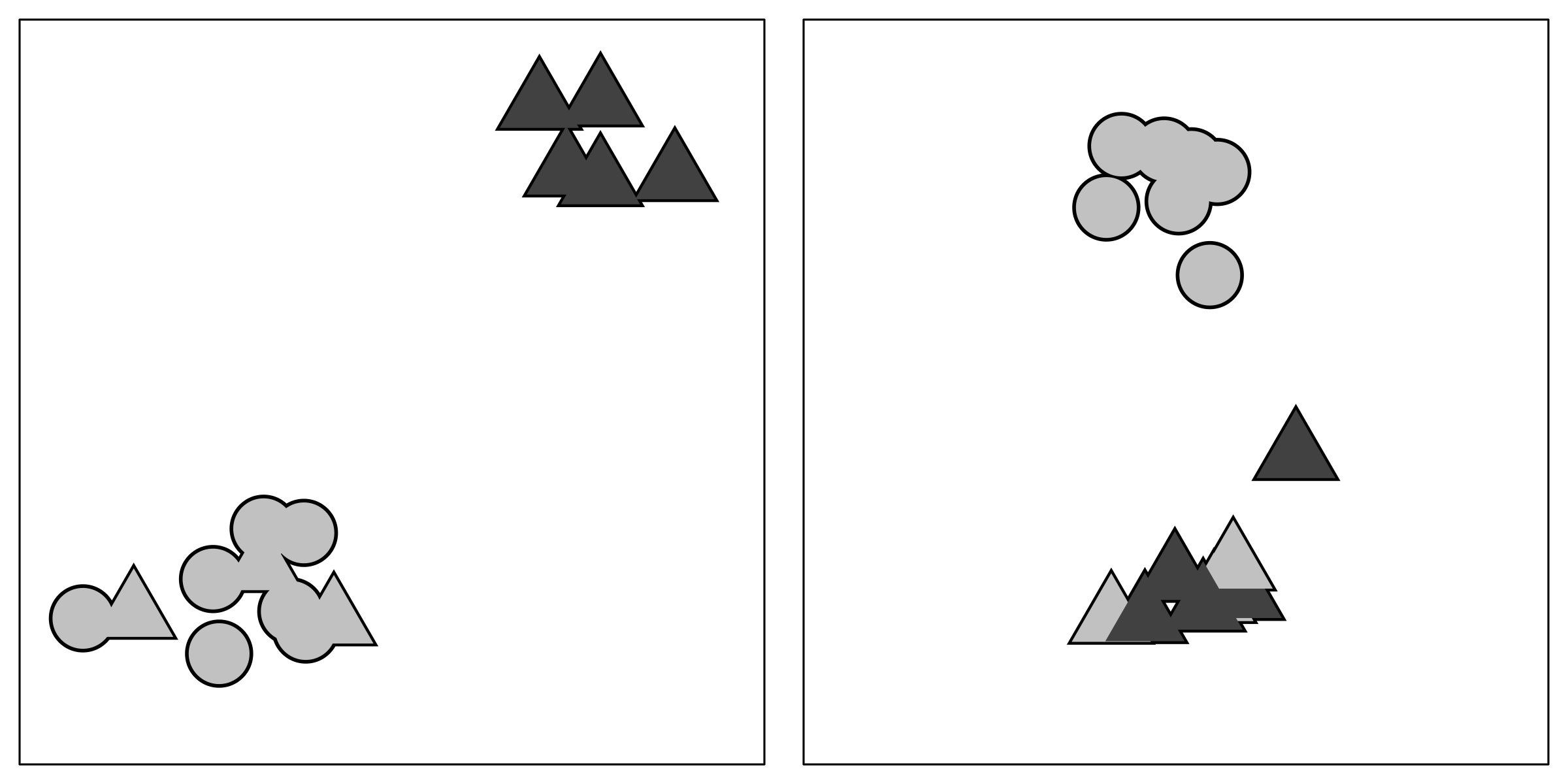}
\caption{\label{fig:clusterviews} Clusters in the first view are represented with dark and light shades of gray, and clusters in the second view are represented with circles and triangles. (i) The clusterings in the two views are independent, i.e. $\Pi$ has rank one, so the shade of gray (dark or light) and shape (circle or triangle) are unassociated. (ii) The clusterings in the two views are the same, i.e. $\Pi$ is diagonal (up to permutation of rows), so the shade of gray (dark or light) and shape (circle or triangle) are perfectly correlated. (iii) The clusterings in the two view are somewhat dependent, i.e. $\Pi$ is neither diagonal nor rank one.  }
\end{figure}

 In Figures~\ref{fig:clusterviews}(i)--(iii), $n=15$ independent pairs $\{(X^{(1)}_i,X^{(2)}_i)\}_{i=1}^n$ are drawn from the model (2.1)--(2.2), for three choices of $\Pi$. The left-hand panel represents the $p_1=2$ features in the first data view, and the right-hand panel represents the $p_2=2$ features in the second data view. For $l = 1, 2$, the observations $\{X^{(l)}_i \}_{i=1}^n$ in the $l$th data view belong to two clusters, where the latent variables $\{Z^{(l)}_i \}_{i=1}^n$ characterize cluster membership in the $l$th data view. Light and dark gray represent the clusters in the first view, and circles and triangles represent the clusters in the second view. 
 
Figures~\ref{fig:clusterviews}(i)--(ii) correspond to two special cases of $\Pi$ that are easily interpretable. In Figure~\ref{fig:clusterviews}(i), $\Pi$ has rank one, i.e. $\Pi = \pi^{(1)} [\pi^{(2)}]^T$, so that the clusterings in the two data views are independent. Thus, whether an observation is light or dark appears to be roughly independent of whether it is a circle or a triangle. In Figure~\ref{fig:clusterviews}(ii), $K^{(1)}=K^{(2)}$ and $\Pi$ is diagonal (up to a permutation of the rows), so that the 
clusterings in the two data views are identical. Thus, all of the circles are light and all of the triangles are dark. Another special case is when $\Pi$ is block diagonal (up to a permutation) with $K_B$ blocks. Then, the clusterings of the two data views agree about the presence of $K_B$ ``meta-clusters" in the data. For example, one clustering might be a refinement of the other, or if one view has clusters $A, B, C, D$, and the other has clusters $a, b, c, d$, it could be that $A\cup B = a$ and $C = b\cup c$ and $D = d$.   

In general, $\Pi$ will be neither exactly rank one nor exactly (block) diagonal; Figure~\ref{fig:clusterviews}(iii) provides such an example. Furthermore, $\hat \Pi$ (an estimator for $\Pi$) almost certainly will be neither.  Nonetheless, examination of $\hat \Pi$ can provide insight into the relationships between the two clusterings. For example, if $\hat \Pi$ is far from rank one, then this suggests that the clusterings in the two data views may be dependent. We will formalize this intuition in Section~\ref{sec:ind}. 

\subsection{Estimation}
\label{sec:fitmodel} 
\subsubsection{Estimation Procedure and Algorithm} 
\label{sec:estimparam} 
Given $n$ independent pairs
$(X^{(1)}_1,X^{(2)}_1),\ldots,(X^{(1)}_n,X^{(2)}_n)$ drawn from the model
\eqref{eq:mixtureMod}--\eqref{eq:defPi}, the log-likelihood takes the form 
 \begin{align} 
\label{eq:jointloglik}
\ell(\theta^{(1)},\theta^{(2)},\pi^{(1)}, \pi^{(2)}, C)=\sum_{i=1}^n\log f(X^{(1)}_i,X^{(2)}_i; \theta^{(1)}, \theta^{(2)}, \pi^{(1)}, \pi^{(2)}, C), 
\end{align} 
where $f(\cdot, \cdot; \theta^{(1)}, \theta^{(2)}, \pi^{(1)}, \pi^{(2)}, C)$ is defined in \eqref{eq:jointdensityC}. A custom expectation-maximization (EM; \citealt{dempster1977maximum, mclachlan2007algorithm}) algorithm could be developed to solve \eqref{eq:jointloglik} for a local optimum (a global optimum is typically unattainable, as \eqref{eq:jointloglik} is non-concave). We instead take a simpler approach.
 Proposition \ref{prop:marginals}
 implies that for $l = 1, 2$, we can estimate $\theta^{(l)}$ and $\pi^{(l)}$ by maximizing the marginal likelihood for the $l$th data view, given by
  \begin{align} 
\label{eq:marginalloglik}
\ell(\theta^{(l)},\pi^{(l)})=\sum_{i=1}^n\log f(X^{(l)}_i; \theta^{(l)}, \pi^{(l)}),
\end{align} 
where $f(\cdot; \theta^{(l)}, \pi^{(l)})$ is defined in \eqref{eq:marginaldensities}. 
Each of these maximizations can be performed using standard EM-based software for model-based clustering of a single data view.  
Let $\hat \theta^{(1)}, \hat \pi^{(1)}, \hat \theta^{(2)},$ and $\hat \pi^{(2)}$ denote the maximizers of \eqref{eq:marginalloglik}. Next, to estimate $C$, we maximize the joint log-likelihood \eqref{eq:jointloglik} evaluated at  $\hat \theta^{(1)}, \hat \pi^{(1)}, \hat \theta^{(2)},$ and $\hat \pi^{(2)}$, 
subject to the constraints imposed by 
Proposition \ref{prop:reparam}:
\begin{align} 
\label{eq:Copt}
\hat{C} &\equiv  \underset{C \in \mathcal{C}_{\hat \pi^{(1)}, \hat \pi^{(2)}}}{\arg \min} \left [ - \ell(\hat \theta^{(1)}, \hat \theta^{(2)}, \hat \pi^{(1)}, \hat \pi^{(2)}, C) \right ],
\end{align} 
where $\mathcal{C}_{\hat \pi^{(1)}, \hat \pi^{(2)}} =  \{ C \in \mathbb R^{K^{(1)} \times K^{(2)}} : C_{kk'} \geq 0, ~ C \hat \pi^{(2)} = 1_{K^{(1)}}, ~ C^T \hat \pi^{(1)} = 1_{K^{(2)}}\}$. Equation \ref{eq:Copt} is a convex optimization problem, which we
solve using a combination of exponentiated gradient descent
\citep{kivinen1997exponentiated} and the Sinkhorn-Knopp algorithm \citep{franklin1989scaling}, as detailed in Appendix \ref{sec:eg} of the Supplementary Materials. 
Details of our approach for fitting the model \eqref{eq:mixtureMod}--\eqref{eq:defPi} are given in Algorithm \ref{alg:estparam}.
\begin{algorithm}
\caption{\label{alg:estparam} Procedure for fitting the model \eqref{eq:mixtureMod}--\eqref{eq:defPi}}
  \begin{enumerate}
  \item Maximize the marginal likelihoods \eqref{eq:marginalloglik} in order to obtain the marginal MLEs  $\hat \theta^{(1)}$, $\hat \pi^{(1)}$ and $\hat \theta^{(2)}$, $\hat \pi^{(2)}$.  This can be done using standard software for model-based clustering.
\item Define matrices $\hat \phi^{(1)}\in\mathbb R^{n\times
K^{(1)}}$ and  $\hat \phi^{(2)}\in\mathbb R^{n\times
K^{(2)}}$ with elements
\begin{align}
  \hat \phi^{(1)}_{ik}= \phi^{(1)}\left(X^{(1)}_i ; \hat \theta^{(1)}_k \right)
  \quad\text{and}\quad \hat \phi^{(2)}_{ik'}= \phi^{(2)}\left(X^{(2)}_i ;
    \hat \theta^{(2)}_{k'} \right).\label{eq:hphi}
\end{align}   
\item Fix a step size $s>0$. Theorem 5.3 from \citet{kivinen1997exponentiated} gives conditions on $s$ that guarantee convergence. 
  \item Let $\hat{C}^1 = 1_{K^{(1)}} 1_{K^{(2)}}^T$.  For $t=1,2,\ldots$ until
    convergence: 
    \begin{enumerate} 
    \item Define $M_{kk'} =   \hat{C}_{kk'}^t  \exp\{sG_{kk'} - 1\},$ where $G_{kk'}=\sum_{i=1}^n\frac{\hat \phi^{(1)}_{ik}\hat \phi^{(2)}_{ik'}}{[\hat \phi^{(1)}_i]^T \mathrm{diag}(\hat \pi^{(1)}) \hat{C}^t \mathrm{diag} (\hat \pi^{(2)}) \hat \phi^{(2)}_i}.$
\item  Let $u^0 = 1_{K^{(2)}}$ and $v^0 = 1_{K^{(1)}}$. For $t' = 1, 2, \ldots$, until convergence:
\begin{enumerate} 
\item $u^{t'} = \frac{1_{K^{(2)}}}{M^T \mathrm{diag}(\hat \pi^{(1)}) v^{t' - 1}}$, \quad $v^{t'} = \frac{1_{K^{(1)}}}{M \mathrm{diag}(\hat \pi^{(2)}) u^{t'}}$, 
\end{enumerate} 
where the fractions denote element-wise vector division.
\item Let $u$ and $v$ be the vectors to which $u^{t'}$ and $v^{t'}$ converge. Let $\hat{C}^{t+1}_{kk'}=  u_k M_{kk'} v_{k'}. $
  \end{enumerate}
  \item Let $\hat{C}$ denote the matrix to which $\hat{C}^t$ converges, and let $\hat\Pi = \mathrm{diag}{(\hat \pi^{(1)})} \hat{C} \mathrm{diag}{(\hat \pi^{(2)})}.$
  \end{enumerate}
\end{algorithm}

\subsubsection{Justification of Estimation Procedure} 
\label{sec:justifyalg} 
The estimation procedure in Section \ref{sec:estimparam} does not maximize the joint likelihood \eqref{eq:jointloglik}; nonetheless, we will argue that it is an attractive approach. 

To begin, in Step 1 of Algorithm \ref{alg:estparam}, we estimate $\theta^{(1)}$ and $\pi^{(1)}$ by maximizing the marginal likelihood \eqref{eq:marginalloglik}. This decision leads to computational advantages, as it enables us to make use of efficient software for clustering a single data view, such as the \verb+mclust+ package \citep{scrucca2016mclust} in \verb+R+.  We can further justify this decision using conditional inference theory. Equation 3.6 in \citet{reid1995roles} extends the definition of ancillary statistics to a setting with nuisance parameters. We show that $X^{(2)}$ is ancillary (in the extended sense of \citealt{reid1995roles}) for $\theta^{(1)}, \pi^{(1)}$, and $C$ by using the definition of conditional densities, and Proposition \ref{prop:marginals}, to rewrite \eqref{eq:jointdensityC} as
\begin{align*} 
f(X^{(1)}, X^{(2)}; \theta^{(1)}, \theta^{(2)}, \pi^{(1)}, \pi^{(2)}, C) =f(X^{(1)} \mid X^{(2)}; \theta^{(1)}, \theta^{(2)}, \pi^{(1)}, \pi^{(2)}, C) f(X^{(2)}; \theta^{(2)}, \pi^{(2)}).
\end{align*}
Thus, \citet{reid1995roles} argues that we should use only $X^{(2)}$, and not $X^{(1)}$, to estimate $\theta^{(2)}$ and $\pi^{(2)}$. In Step 1 of Algorithm \ref{alg:estparam}, we are doing exactly this. 

In Steps 3--5 of Algorithm \ref{alg:estparam}, we maximize $\ell(\hat \theta^{(1)},\hat \theta^{(2)},\hat \pi^{(1)}, \hat \pi^{(2)}, \cdot)$, giving $\hat C$, which is a pseudo maximum likelihood estimator for $C$ in the sense of \citet{gong1981pseudo}. This decision also leads to computational advantages, as it enables us to make use of efficient convex optimization algorithms in estimating $C$. Results in \citet{gong1981pseudo} suggest that when $\hat \theta^{(1)}$, $\hat \theta^{(2)}$, $\hat \pi^{(1)}$, and $\hat \pi^{(2)}$ are good estimates, $\hat C$ is so as well.

\subsection{Selection of the Number of Clusters} 
\label{sec:tun}
In Sections 2 and 3, our discussion assumed that $K^{(1)}$ and $K^{(2)}$ are known. However, this is rarely the case in practice.
    Recall that we estimate $\theta^{(1)}$ and $\pi^{(1)}$ by maximizing the marginal likelihood \eqref{eq:marginalloglik}, which amounts to performing model-based clustering of $X^{(1)}$ only. Thus, to select the number of clusters $K^{(1)}$, we can make use of an extensive literature (reviewed in e.g. \citealt{mirkin2011choosing}) on choosing the number of clusters when clustering a single data view. For example, we can use AIC or BIC to select $K^{(1)}$ and $K^{(2)}$. 

\section{Testing Whether Two Clusterings are Independent} 
\label{sec:ind}

\subsection{A Brief Review of Pseudo Likelihood Ratio Tests} 
\label{sec:plrt}
Let $\ell(\alpha, \beta, \gamma)$ be the log-likelihood function for a random sample, where $\mathcal{A}$ is the parameter space of $\alpha$.  Given a null hypothesis $H_0: \alpha = \alpha_0$ for some $\alpha_0 \in \mathcal{A}$, an alternative hypothesis $H_1: \alpha \neq \alpha_{0}$, and an estimator $\hat \gamma$, the pseudo likelihood ratio statistic \citep{self1987asymptotic} is defined to be $\log \tilde{\Lambda} \equiv \underset{\alpha, \beta}{\sup} ~\ell(\alpha, \beta, \hat{\gamma}) -  \underset{\beta}{\sup} ~ \ell(\alpha_0, \beta, \hat{\gamma} )$. Let $\alpha^*$ be the true parameter value for $\alpha$. If  $\alpha^*$ is an interior point of $\mathcal{A}$, then under some regularity conditions, if $H_0$ holds, then 
$2 \log \tilde{\Lambda} \overset{d}{\longrightarrow}  \chi^2_r,$
where $r$ is the dimension of $\mathcal{A}$ \citep{chen2010asymptotic}. 

\subsection{A Pseudo Likelihood Ratio Test for Independence} 
\label{sec:testind} 
In this subsection, we develop a test for the null hypothesis that 
 $H_0: C = 1_{K^{(1)}} 1_{K^{(2)}}^T$, or equivalently, that $H_0: \Pi = \pi^{(1)} (\pi^{(2)})^T$: that is, we test whether $Z^{(1)}$ and $Z^{(2)}$ are independent, i.e. whether the cluster memberships in the two data views are independent.  We could use a likelihood ratio test statistic to test $H_0$,
\begin{align} 
\label{eq:likstatindep}
\log \Lambda &\equiv \operatornamewithlimits{sup}_{\theta^{(1)},\theta^{(2)},\pi^{(1)}, \pi^{(2)}, C}  \ell(\theta^{(1)},\theta^{(2)},\pi^{(1)}, \pi^{(2)}, C)- \operatornamewithlimits{sup}_{\theta^{(1)}, \theta^{(2)}, \pi^{(1)}, \pi^{(2)}} \ell(\theta^{(1)},\theta^{(2)},  \pi^{(1)}, \pi^{(2)}, 1_{K^{(1)}} 1_{K^{(2)}}^T) \nonumber \\  
&= \operatornamewithlimits{sup}_{\theta^{(1)},\theta^{(2)},\pi^{(1)}, \pi^{(2)}, C}  \ell(\theta^{(1)},\theta^{(2)},\pi^{(1)}, \pi^{(2)}, C) - [\ell(\hat \theta^{(1)},\hat \pi^{(1)})  + \ell(\hat \theta^{(2)}, \hat \pi^{(2)})],
\end{align} 
where the second equality follows from noticing that substituting  $C = 1_{K^{(1)}} 1_{K^{(2)}}^T$ into \eqref{eq:jointloglik} yields
\begin{align} 
\label{eq:loglik-rank1}
\ell(\theta^{(1)}, \theta^{(2)}, \pi^{(1)}, \pi^{(2)}, C) = \ell(\theta^{(1)}, \pi^{(1)}) + \ell(\theta^{(2)}, \pi^{(2)}), 
\end{align} 
where $\ell(\theta^{(l)}, \pi^{(l)})$ for $l = 1, 2$ are defined in \eqref{eq:marginalloglik}, and recalling the definition of $\hat \theta^{(1)}$,  $\hat \pi^{(1)}$, $\hat \theta^{(2)}$, and $\hat \pi^{(2)}$ as the maximizers of \eqref{eq:marginalloglik}. However, \eqref{eq:likstatindep} requires maximizing $\ell(\theta^{(1)},\theta^{(2)},\pi^{(1)}, \pi^{(2)}, C)$, which would require a custom EM algorithm; furthermore, the resulting test statistic will typically involve the difference between two local maxima (since each term in \eqref{eq:likstatindep} requires fitting an EM algorithm). This leads to erratic behavior, such as negative values of $\log\Lambda$. 

Therefore, instead of taking the approach in
 \eqref{eq:likstatindep}, we develop a pseudo likelihood ratio test, as in
 Section~\ref{sec:plrt}. We use the marginal MLEs, $\hat \theta^{(1)},  \hat \pi^{(1)}$, and $\hat \theta^{(2)}, \hat \pi^{(2)}$, instead of performing the joint optimization in \eqref{eq:likstatindep}. This leads to the 
test statistic
\begin{align} 
\log \tilde{\Lambda} &\equiv \operatornamewithlimits{sup}_{C \in \mathcal{C}_{\hat \pi^{(1)}, \hat \pi^{(2)}}} \ell( \hat \theta^{(1)}, \hat \theta^{(2)}, \hat \pi^{(1)}, \hat \pi^{(2)}, C)  - \ell(  \hat \theta^{(1)}, \hat \theta^{(2)}, \hat \pi^{(1)}, \hat \pi^{(2)}, 1_{K^{(1)}} 1_{K^{(2)}}^T) \nonumber \\ 
&= \ell (\hat \theta^{(1)}, \hat \theta^{(2)}, \hat \pi^{(1)}, \hat \pi^{(2)}, \hat{C}) - [\ell(\hat \theta^{(1)}, \hat \pi^{(1)}) + \ell(\hat \theta^{(2)}, \hat \pi^{(2)}) ] \label{eq:pseudolikstatindep}\\ 
&= \sum_{i=1}^n\log\left[\frac{(\hat \phi^{(1)}_i)^T\mathrm{diag}(\hat \pi^{(1)}) \hat{C} \mathrm{diag}(\hat \pi^{(2)})\hat \phi^{(2)}_i}{(\hat \phi^{(1)}_i)^T \hat \pi^{(1)} (\hat \pi^{(2)})^T\hat \phi^{(2)}_i}\right]\label{eq:lambda}.
\end{align}
where $\hat{C}$ in \eqref{eq:pseudolikstatindep} is defined in \eqref{eq:Copt}, $\mathcal{C}_{\hat \pi^{(1)}, \hat \pi^{(2)}}$ is defined in Proposition~\ref{prop:reparam}, $\hat \phi^{(1)}$ and $\hat \phi^{(2)}$ are defined in \eqref{eq:hphi}, and the last equality follows from \eqref{eq:jointloglik}, \eqref{eq:marginalloglik}, and \eqref{eq:hphi}. In addition to taking advantage of the computationally efficient estimation procedure described in Section \ref{sec:estimparam}, the pseudo likelihood ratio test statistic does not exhibit the erratic behavior exhibited by the likelihood ratio test statistic. This stability comes from all three terms in \eqref{eq:pseudolikstatindep} involving the same local maxima (as opposed to different local maxima). 

\subsection{Approximating the Null Distribution of $\log \tilde{\Lambda}$}
\label{sec:approxnull}
The discussion in Section~\ref{sec:plrt} suggests that under $H_0: C = 1_{K^{(1)}} 1_{K^{(2)}}^T$, one might expect that $2 \log \tilde{\Lambda} \overset{d}{\longrightarrow} \chi^2_r,$
where $r = (K^{(1)} - 1)(K^{(2)} - 1)$ is the dimension of $\mathcal{C}_{\pi^{(1)}, \pi^{(2)}}$. However, this approximation performs poorly in practice, due to violations of the regularity conditions in \citet{chen2010asymptotic}. Furthermore, we will often be interested in data applications in which $n$ is relatively small. Hence, we propose a permutation approach. We observe from \eqref{eq:loglik-rank1} that under $H_0$, the log-likelihood is identical under any permutation of the order of the samples in each view. Hence, we take $B$ random permutations of the samples $X^{(2)}_i$ from the second view, and compare the observed value of $\log \tilde{\Lambda}$ to its empirical distribution in these permutation samples. Details are given in Algorithm~\ref{alg:permuteind}. Since $\hat \phi^{(1)}$, $\hat \phi^{(2)}$, $\hat \pi^{(1)}$, and $\hat \pi^{(2)}$ are invariant to permutation, for each permutation we need only to estimate $C$. This is another advantage of our test over the likelihood ratio test discussed in Section \ref{sec:testind}, which would require repeating the EM algorithm in every permutation.  
\begin{algorithm}
\caption{A Permutation Approach for Testing $H_0: C = 1_{K^{(1)}} 1_{K^{(2)}}^T $} \label{alg:permuteind}
  \begin{enumerate}
  \item Compute $\log \tilde{\Lambda}$ according to \eqref{eq:lambda} using the original data, $X^{(1)}$ and $X^{(2)}$.  
  \item For $b=1,\ldots,B$, where $B$ is the number of permutations:
  \begin{enumerate}
  \item Permute the observations in $X^{(2)}$ to obtain $X^{(2, *b)}$.
  \item  Compute $ \log \tilde{\Lambda}^{*b}$ according to \eqref{eq:lambda} based on $X^{(1)}$ and $X^{(2, *b)}$.
  \end{enumerate}
  \item The p-value for testing $H_0: C = 1_{K^{(1)}} 1_{K^{(2)}}^T $ is given by 
  $\frac{1}{B} \sum_{b=1}^B 1_{ \left\{ \log \tilde{\Lambda} \leq  \log \tilde{\Lambda}^{*b}  \right\} }.$
\end{enumerate}
\end{algorithm}
Even when we reject the null hypothesis, the clusters could be only weakly dependent; thus, it is helpful to measure the strength of association between the views. Recalling from Section \ref{sec:interpretpi} that $\text{rank}(\Pi) = 1$ implies independence of the clusterings in the two data views, we propose to calculate the effective rank \citep{vershynin_2012} of $\hat \Pi$, defined in Algorithm \ref{alg:estparam} --  the ratio of the sum of the singular values of $\hat \Pi$, and the largest singular value of $\hat \Pi$. The effective rank of a matrix is bounded between 1 and its rank, and the matrix is far from rank-1 when its effective rank is far from 1. For example, in Figure \ref{fig:clusterviews}(iii), the effective rank of $\Pi$ is 1.5, and is upper bounded by 2. Thus, the effective rank of $\hat \Pi$ is bounded between 1 and $\min \{K^{(1)}, K^{(2)} \}$, and $\hat \Pi$ is far from rank-1 when its effective rank is far from 1. 

\section{Simulation Results}
\label{sec:sim}
To investigate the Type I error and power of our test,  
we generate data from \eqref{eq:mixtureMod}--\eqref{eq:defPi}, with 
\begin{align} 
\label{eq:simPi}
\Pi = \frac{1-\delta}{K^2} 1_K 1_K^T + \frac{\delta}{K} I_K,
\end{align} for $K = 6$ and for a range of values of $\delta \in [0,1]$, where $\delta = 0$ corresponds to independent clusterings, and $\delta = 1$ corresponds to identical clusterings. We draw the observations in the $l$th data view from a Gaussian mixture model, for which the $k$th mixture component is a $N_p(\mu^{(l)}_k, \Sigma^{(l)})$ distribution, with $p = 10$, and with $\mu^{(l)}_k$ given in Appendix \ref{sec:meanmat} of the Supplementary Materials. 

\begin{figure}[H]
\centering
\includegraphics[scale=0.55]{{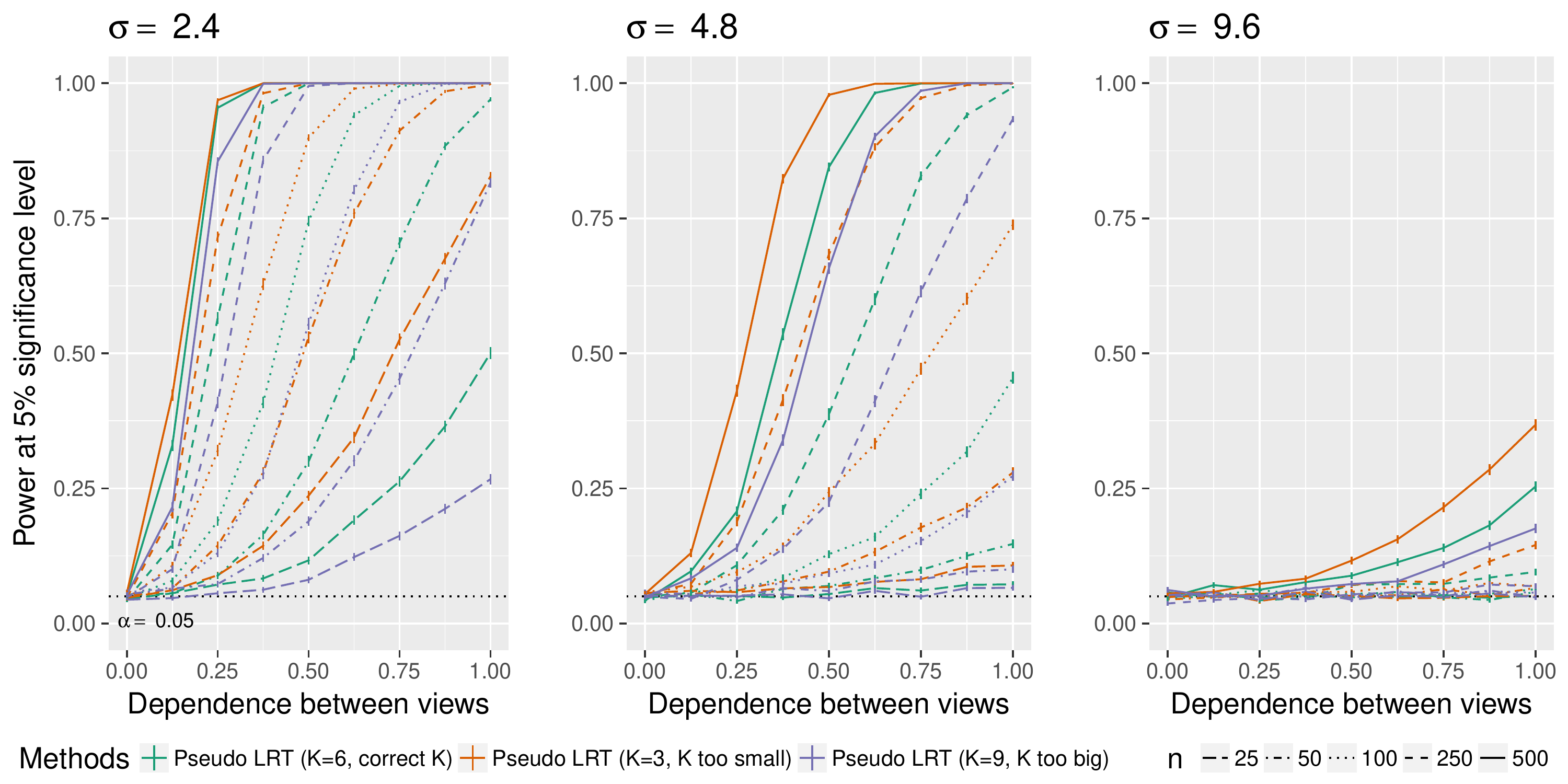}}
\caption{\label{fig:power6} Power of the pseudo likelihood ratio test of $H_0: C = 1_{K_1} 1_{K_2}^T$ with $p = 10$, $K = 6$  and $\sigma \in  \{ 2.4, 4.8, 9.6\}$ in the simulation setting described in Section~\ref{sec:sim}. The $x$-axis displays  $\delta$, defined in \eqref{eq:simPi},  and the $y$-axis displays the power.  }
\end{figure}

We simulate 2000 data sets for $\Sigma^{(1)} = \Sigma^{(2)} = \sigma^2 I_p$ for a range of values of $\sigma$ and $n$, and evaluate the power of the pseudo likelihood ratio test of $H_0: C = 1_{K_1} 1_{K_2}^T$ described in Section \ref{sec:testind} at nominal significance level $\alpha = 0.05$, when the number of clusters is correctly and incorrectly specified.	To perform Step 1 of Algorithm \ref{alg:estparam}, we use the package \verb+mclust+ in \verb+R+ to fit Gaussian mixture models with a common $\sigma^2 I_p$ covariance matrix (the ``EII" covariance structure in \verb+mclust+). We use $B = 200$ permutation samples in Step 2 of Algorithm \ref{alg:permuteind}.  Simulations in this paper were conducted using the \verb+simulator+ package \citep{bien2016simulator} in \verb+R+. Results are shown in Figure \ref{fig:power6}.

The pseudo likelihood ratio test controls the Type I error close to the nominal $\alpha = 0.05$ level, even when the number of clusters is misspecified. Power tends to increase as $\delta$ (defined in \eqref{eq:simPi}) increases, and tends to decrease as $\sigma$ increases. Compared to using the correct number of clusters, using too many clusters yields lower power, but using too few clusters can sometimes yield higher power (e.g. in the middle panel of Figure \ref{fig:power6}). This is because, when the signal-to-noise ratio is low, the true clusters are not accurately estimated; thus, combining several true clusters into a single ``meta-cluster" can sometimes, but not always, lead to improved agreement between clusterings across the two data views. We explore the impact of the choice of $K$ on the performance of the pseudo likelihood ratio test in Appendix \ref{sec:chooseK} of the Supplementary Materials. 

Additional values of $K$ and $p$ are investigated in Appendix \ref{sec:morepower1}.

\section{Connection to the G-test for Independence and Mutual Information} 
\label{sec:meila}

Let $\hat M^{(1)} = (\hat M^{(1)}_1, \ldots, \hat M^{(1)}_n)$ and $\hat M^{(2)} = (\hat M^{(2)}_1, \ldots, \hat M^{(2)}_n)$ denote the results of applying a clustering procedure to $X^{(1)}$ and $X^{(2)}$ respectively. In this notation, $\hat M^{(1)}_i \in \{1, \ldots, K^{(1)}\}$ and $\hat M^{(2)}_i \in \{1, \ldots, K^{(2)}\}$ denote the estimated cluster assignment for the $i$th observation in the two views. To test whether $Z^{(1)}$ and $Z^{(2)}$ are independent, we could naively apply tests on $\hat M^{(1)}$ and $\hat M^{(2)}$ for whether two categorical variables are independent. For instance, we could use the $G$-test statistic for independence (Chapter 3.2, \citealt{agresti2003categorical}), given by 
\begin{align} 
\label{eq:Gstat} 
G^2(\hat M^{(1)}, \hat M^{(2)}) = 2\sum \limits_{k=1}^{K^{(1)}} \sum \limits_{k'=1}^{K^{(2)}} \hat N_{kk'} \log \left[  (n\hat N_{kk'})/(\hat N_{k.}\hat N_{.k'}) \right],
\end{align}
where $\hat N_{kk'} = |\{i \in \{1, \ldots, n\} : \hat M^{(1)}_i = k, \hat M^{(2)}_i = k'\}|$, $\hat N_{.k'} = \sum \limits_{k} \hat N_{kk'}$, and $\hat N_{k.} = \sum \limits_{k'} \hat N_{kk'}$. Under the model
 $\hat N_{kk'}  ~\overset{\mathrm{ind}}{\sim} \text{ Poisson } (\mu_{kk'}),  \log \mu_{kk'} = \alpha_k + \beta_{k'} + \gamma_{kk'},$
 the $G$-test statistic for independence \eqref{eq:Gstat} is a likelihood ratio test statistic for testing the null hypothesis of independence, i.e. for testing $H_0$: $\gamma_{kk'} = 0 \text{ for all } k, k'$. Thus, under $H_0$: $\gamma_{kk'} = 0$, 
\begin{align} 
G^2(\hat M^{(1)}, \hat M^{(2)}) \overset{d}{\rightarrow} \chi^2_{(K^{(1)} - 1)(K^{(2)} -1)}. \label{eq:chisq}
\end{align} 

The $G$-test statistic for independence \eqref{eq:Gstat} relies on an assumption which is violated in our setting, namely that $\{(\hat M^{(1)}_i, \hat M^{(2)}_i)\}_{i=1}^n$ are independent and identically distributed samples from the distribution of $(Z^{(1)}, Z^{(2)})$. It is nonetheless a natural approach to the problem of comparing two views' clusterings. In fact, the mutual information of \citet{meilua2007comparing} for measuring the similarity between two clusterings of a single $n \times p$ dataset can be written as a scaled version of the $G$-test statistic; when applied to instead measure the similarity between $\hat M^{(1)}$ and $\hat M^{(2)}$, the mutual information $I(\hat M^{(1)}, \hat M^{(2)})$ is given by 
\begin{align} 
I(\hat M^{(1)}, \hat M^{(2)}) = G^2(\hat M^{(1)}, \hat M^{(2)})/2n. \label{eq:minf}
\end{align} 
While the proposed pseudo likelihood ratio test statistic \eqref{eq:lambda} for testing independence of $Z^{(1)}$ and $Z^{(2)}$ does not resemble the simple $G$-test statistic for independence in \eqref{eq:Gstat}, we show here that they are in fact quite related. 

Let $\hat r_i^{(1)} = \frac{\hat \phi^{(1)}_i}{1_{K^{(1)}}^T \hat{\phi}^{(1)}_i}$ and $\hat r_i^{(2)} =  \frac{\hat \phi_i^{(2)}}{1_{K^{(2)}}^T \hat{\phi}^{(2)}_i}$ be the vectors giving the soft-clustering assignment weights (or ``responsibilities") for the $i$th observation in the two views, where $\hat\phi_i$ is defined in \eqref{eq:hphi}.  We rewrite the pseudo likelihood ratio test statistic \eqref{eq:lambda} as
\begin{align} 
\log\tilde{\Lambda}\left (\hat \Pi, \{\hat r^{(1)}_i, \hat r^{(2)}_i\}_{i=1}^n \right ) = \sum_{i=1}^n\log\left[\frac{(\hat r_i^{(1)})^T\hat \Pi \hat r_i^{(2)}}{ (\hat r^{(1)}_i )^T(\hat \Pi 1_{K^{(2)}}) (1_{K^{(1)}}^T \hat \Pi) \hat r_i^{(2)}}\right], \label{eq:softstat}
\end{align} 
where $\hat \Pi$ is defined in Algorithm \ref{alg:estparam}. In the following proposition, we consider replacing the ``soft" cluster assignments $\hat r^{(1)}_i$ and $\hat r^{(2)}_i$ with ``hard" cluster assignments, and replacing the estimate $\hat \Pi$ derived from the ``soft" cluster assignments with an estimate derived from ``hard" cluster assignments, in \eqref{eq:softstat}. In what follows, 
\begin{align} 
\label{eq:marginalclustest}
\hat M_i^{(1)} \equiv \underset{k \in \{1, 2, \ldots, K^{(1)} \}}{\text{arg max}} ~ \hat r^{(1)}_{ik},  \;\;\;
\hat M_i^{(2)} \equiv \underset{k' \in \{1, 2, \ldots, K^{(2)} \}}{\text{arg max}} ~ \hat r^{(2)}_{ik'}.
\end{align}
\begin{prop} 
Let $\hat M^{(1)}$ and $\hat M^{(2)}$ be the estimated model-based cluster assignments in each data view defined by \eqref{eq:marginalclustest}. Let $\hat N$ be the matrix with entries $\hat N_{kk'}$ containing the number of observations assigned to cluster $k$ in view 1 and cluster $k'$ in view 2.  Then, 
\begin{align} 
\log\tilde{\Lambda}(\hat N/n, \{e_{\hat M^{(1)}_i}, e_{\hat M^{(2)}_i}\}_{i=1}^n) = nI(\hat M^{(1)}, \hat M^{(2)}) = G^2(\hat M^{(1)}, \hat M^{(2)})/2, \label{eq:meila}
\end{align}  where $\log \tilde \Lambda(\cdot, \cdot)$ is defined in \eqref{eq:softstat}, and  $e_t$ is the unit vector that contains a 1 in the $t$th element.
\label{prop:meilaConnection}
\end{prop}
Proposition \ref{prop:meilaConnection} follows by algebra, and says that replacing the soft cluster assignments in the pseudo likelihood ratio test statistic of Section \ref{sec:ind} with hard cluster assignments yields \emph{exactly} the $G$-test statistic for independence \eqref{eq:Gstat} (and the mutual information given in \eqref{eq:minf})! In fact, in the special case of fitting multiple-view Gaussian mixtures with common covariance matrix $\sigma^2 I_{p_1}$ in the first view and $\sigma^2 I_{p_2}$ in the second view, we will show that as $\sigma \rightarrow 0$, and the soft cluster assignments converge to hard cluster assignments, the pseudo likelihood ratio test statistic converges to the $G$-test for independence. In what follows, $\log \tilde \Lambda \equiv \log \tilde \Lambda\left (\hat \Pi, \{\hat r^{(1)}_i, \hat r^{(2)}_i\}_{i=1}^n \right )$, as in \eqref{eq:lambda} and \eqref{eq:softstat}.
\begin{prop}
\label{prop:connectmeila}
Let $\sigma^2 > 0$. Suppose that to compute $\log \tilde \Lambda$, we fit the model \eqref{eq:mixtureMod}--\eqref{eq:defPi}, for $\phi^{(1)}$ and  $\phi^{(2)}$ densities of Gaussian distributions with covariance matrices $\sigma^2 I_{p_1}$ and $\sigma^2 I_{p_2}$ respectively. Let $\tilde M^{(1)}$ and $\tilde M^{(2)}$ denote the results of applying k-means clustering on the two data views. Then, as $\sigma^2 \rightarrow 0$, $\log \tilde{\Lambda} \rightarrow  nI(\tilde M^{(1)}, \tilde M^{(2)}) = G^2(\tilde M^{(1)},  \tilde M^{(2)})/2$.
\end{prop} 
Proposition \ref{prop:connectmeila} is proven in Appendix \ref{sec:proofconnectmeila} of the Supplementary Materials. When $\sigma^2 > 0$, the pseudo likelihood ratio test statistic, the $G$-test statistic, and the mutual information are not equivalent. We can thus think of the pseudo likelihood ratio test statistic as reflecting the uncertainty associated with the clusterings obtained on the two views, and the $G$-test statistic and the mutual information as ignoring the uncertainty associated with the clusterings. This suggests that the pseudo likelihood ratio test of Section \ref{sec:testind} outperforms the $G$-test for independence when the sample size is small and/or there is little separation between the clusters. 

To confirm this intuition, we return to the simulation set-up described in Section~\ref{sec:sim}, and compare the performances of the pseudo likelihood ratio test \eqref{eq:lambda} and the G-test for independence \eqref{eq:Gstat} for testing $H_0: C=1_{K^{(1)}} 1_{K^{(2)}}^T$. We obtain p-values for \eqref{eq:Gstat} using the $\chi^2$ approximation from \eqref{eq:chisq}, and using a permutation approach, where we take $B$ permutations of the elements of $\hat M^{(2)}$, and compare the observed value of \eqref{eq:Gstat} to its empirical distribution in these permutation samples. The results are shown in Figure \ref{fig:meilaK6}; we see that the two tests yield similar power when the sample size is larger and/or the value of $\sigma$ is smaller, and that the pseudo likelihood ratio test yields higher power than the G-test for independence when the sample size is smaller and/or the value of $\sigma$ is larger. We note that the $\chi^2$ approximation for the G-test from \eqref{eq:chisq} does not control the Type I error. Additional values of $p$ and $K$, additional values of $\Sigma^{(l)}$, and non-Gaussian finite mixture models are investigated in Appendices \ref{sec:moreK}, \ref{sec:morecov}, and \ref{sec:misspec} of the Supplementary Materials, respectively; the results are similar to those described in this section. 

\section{Application to the Pioneer 100 Wellness Project \citep{price2017wellness}}
\label{sec:app}
\subsection{Introduction to the Scientific Problem}
\label{sec:pioneer100} 
In the P100 Wellness Project \citep{price2017wellness}, multiple biological data types were collected at multiple timepoints for 108 healthy participants. For each participant, whole genome sequences were measured, activity tracking data were collected daily over nine months, and clinical laboratory tests, metabolomes, proteomes, and microbiomes were measured at three-month, six-month, and nine-month timepoints. The P100 study aims to optimize wellness of the participants through personalized healthcare recommendations. In particular, clinical biomarkers measured at baseline were used to make personalized health recommendations. 

As an alternative approach, we could identify subgroups of individuals with similar clinical profiles using cluster analysis, and then develop interventions tailored to each subgroup. It is tempting to identify these subgroups using not just clinical data at baseline, but also other types of data (e.g. proteomic data) at other timepoints. We could do this by applying a multi-view consensus clustering method (e.g.  \citealt{shen2009integrative}). However, such an approach assumes that there is a single true clustering underlying all data types at all timepoints. Therefore, before applying a consensus clustering approach, we should determine whether there is any evidence that the clusterings underlying the data types and/or timepoints are at all related (in which case consensus clustering may lead to improved estimation of the clusters) or whether the clusterings are completely unrelated (in which case one would be better off simply performing a separate clustering of the observations in each view). In what follows, we will use the hypothesis test developed in Section~\ref{sec:ind} to determine whether clusterings of P100 participants based on clinical, proteomic, and genomic data are dependent across timepoints, and across data types. 

\subsection{Data Analysis} 
At each of the three timepoints, 207 clinical measurements, 268 proteomic measurements, and 642 metabolomic measurements were available for $n=108$ observations. In the following, we define a data view to be a single data type at a single timepoint. In each view, we removed features missing in more than 25\% of participants, and removed participants missing more than 25\% of features. Next, features in each view with standard deviation 0 were removed. The remaining missing data were imputed using nearest neighbors imputation in the \verb=impute= package in \verb=R= \citep{impute}. Features in each view were then adjusted for gender using linear regression.  Finally, the remaining features were scaled to have standard deviation 1. As in Section~\ref{sec:sim}, we consider the model \eqref{eq:mixtureMod}--\eqref{eq:defPi} 
under the assumption that each component in the mixture is drawn from a Gaussian distribution. For each data view, we fit the model using the \verb+mclust+ package in \verb+R+, with a common $\sigma^2 I$ covariance matrix (the ``EII" covariance structure in \verb+mclust+). To test $H_0: C = 1_{K^{(1)}} 1_{K^{(2)}}^T$, we compute p-values using the permutation approximation discussed in Section~\ref{sec:approxnull} with $B = 10^5$. Based on the results in Appendix \ref{sec:chooseK} of the Supplementary Materials, we choose the number of clusters in each view by BIC under the constraint that the number of clusters is greater than one.

\begin{figure}[H] 
\noindent (i) \hspace{77mm} (ii) \\ 
\includegraphics[scale=0.25]{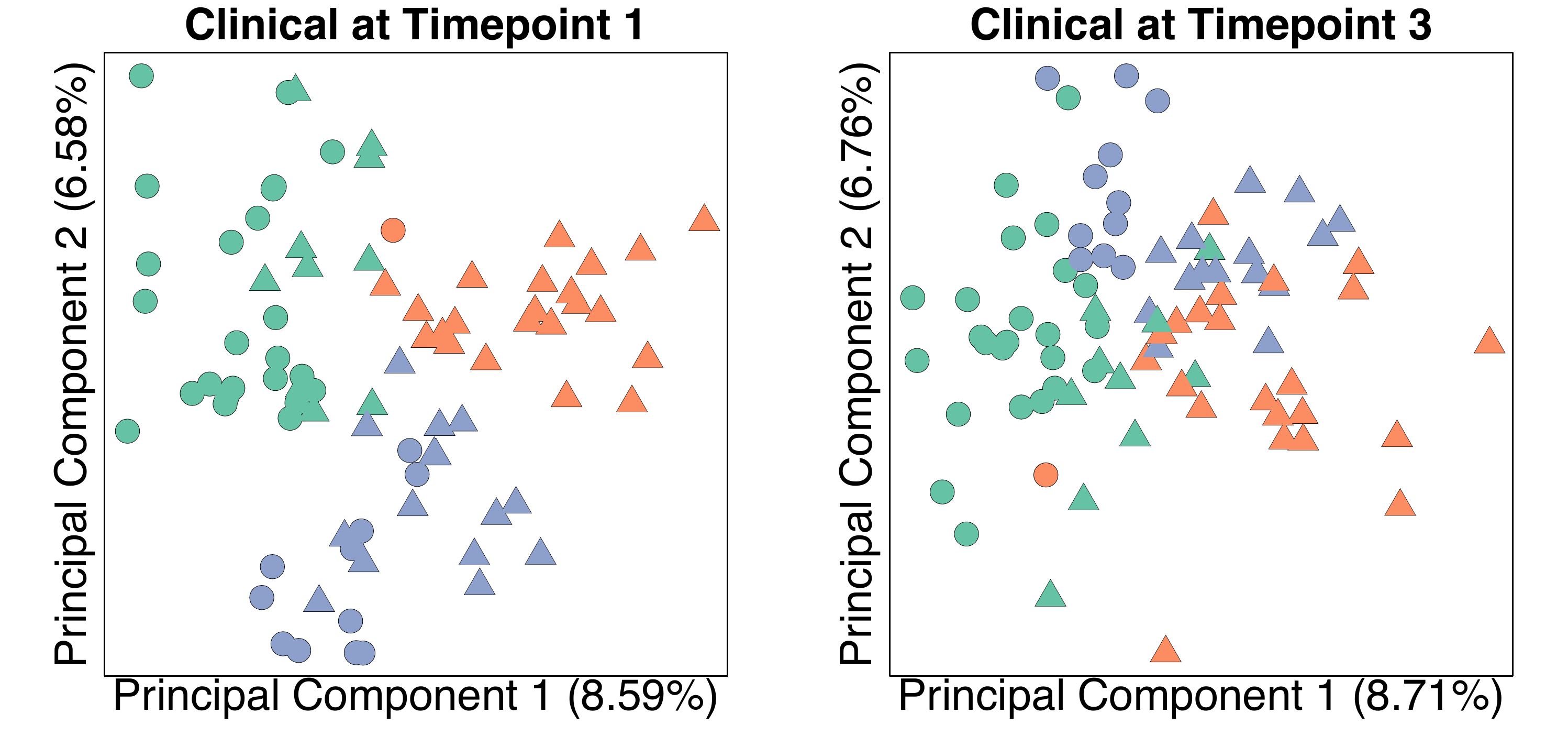} \hspace{10mm} 
\includegraphics[scale=0.25]{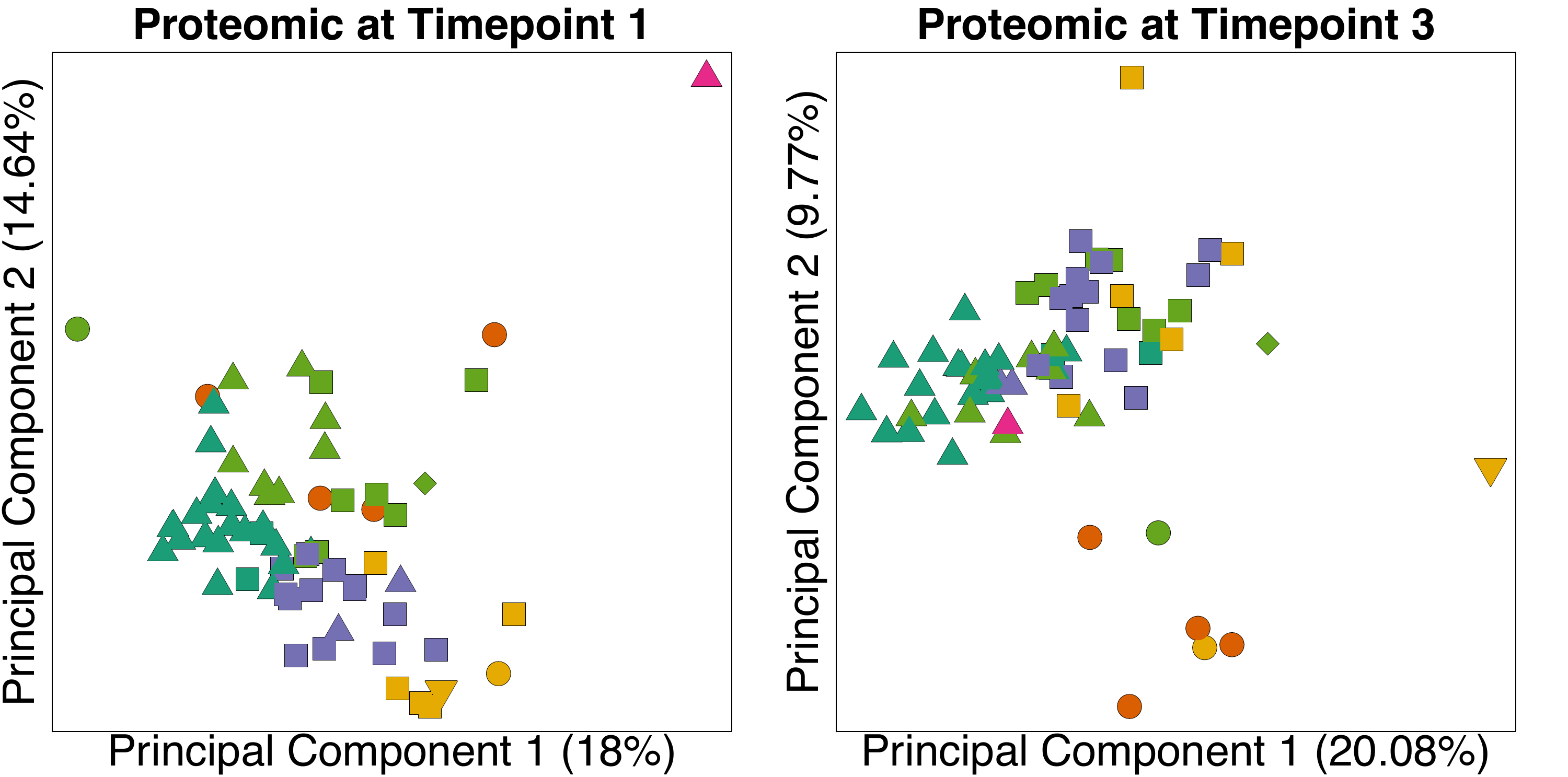} \\
(iii) \\
\includegraphics[scale=0.25]{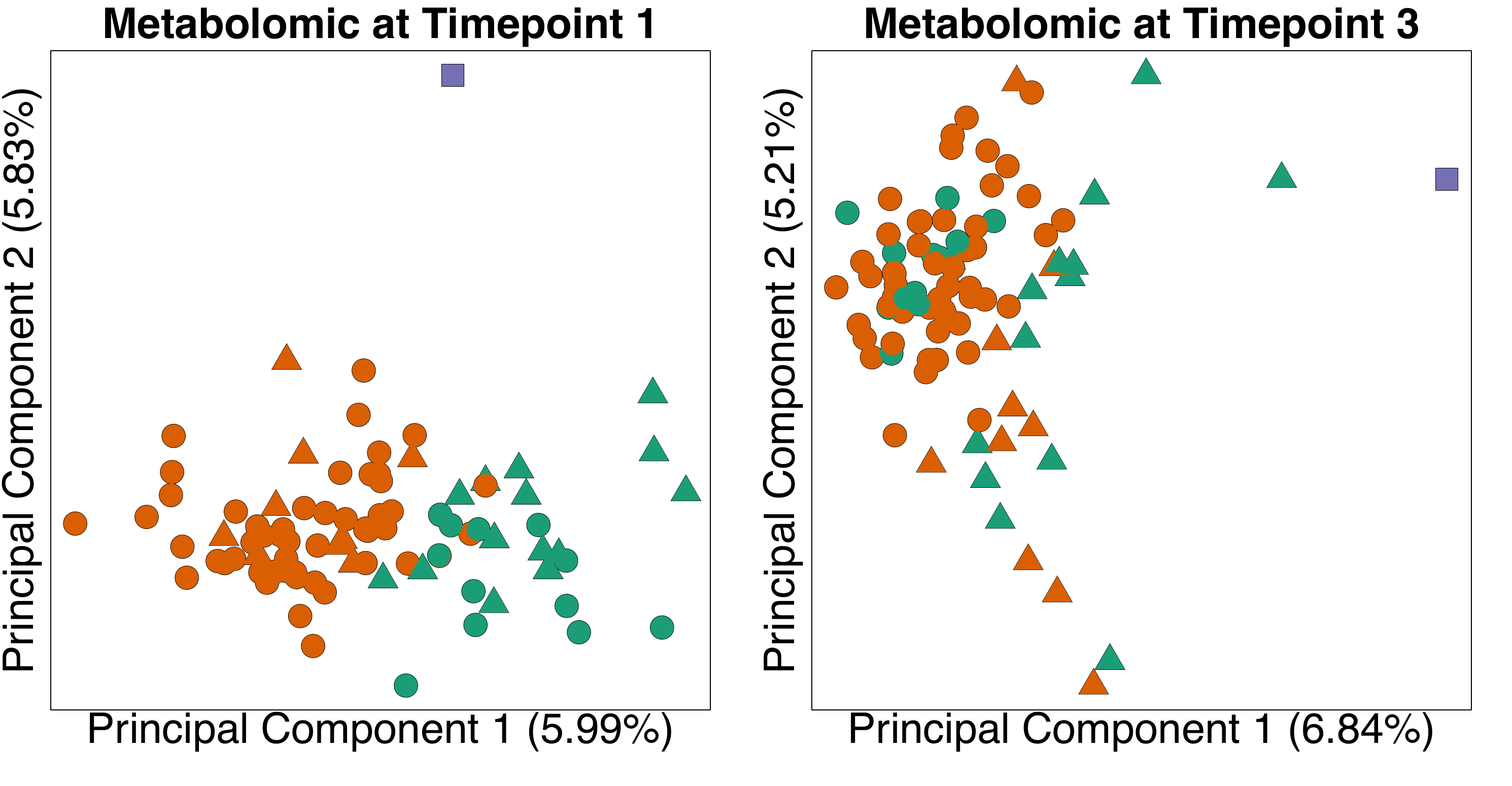}
\caption{\label{fig:clusters} For three different data types, a comparison of the clustering at the first timepoint (represented with colors) with the clustering at the third timepoint (represented with shapes).  In each data type, there is strong evidence of dependence (p-value $< 0.0001$).  The data types are (i) clinical measurements, (ii) proteomic measurements, and (iii) metabolomic measurements.}
\end{figure}

We now compare the clusterings in the clinical data at the first and third timepoints, the clustering in the proteomic data at the first and third timepoints, and the clusterings in the metabolomic data at the first and third timepoints. The sample sizes and results are reported in Table~\ref{tab:pval}. For each data type, the clusters found at each timepoint are displayed in Figure \ref{fig:clusters}.  

We find strong evidence that for each data type, the clusterings at the first and third timepoints are not independent. We further measure the strength of dependence through the effective rank of $\hat \Pi$, as described in Section \ref{sec:approxnull}. For the clusterings in the clinical data, the effective rank of $\hat \Pi$ is 1.63, and is upper bounded by 2. For the clusterings in the proteomic data, the effective rank of $\hat \Pi$ is 1.90, and is upper bounded by 5. For the clusterings in the metabolomic data, the effective rank of $\hat \Pi$ is 1.2, and is upper bounded by 3. These results  suggest that the strengths of association for the clusterings estimated on the clinical data, the proteomic data, and the metabolomic data, are strong, moderate, and weak respectively. The fact that the clusterings estimated on some data types are strongly dependent over time provides evidence that they are scientifically meaningful. Furthermore, it suggests that performing consensus clustering on some data types (e.g. clinical data and proteomic data) across timepoints may be reasonable.

We now focus on comparing clusterings in the clinical, proteomic, and metabolomic data at a single timepoint. The sample sizes and results are reported in Table~\ref{tab:pval}. 

The results provide modest evidence that proteomic and metabolomic data at a given timepoint are dependent, and provide weak evidence that clinical and metabolomic data are dependent. However, on balance, the evidence that the clusterings are dependent across data types is weaker than we might expect. This suggests to us that the underlying subgroups defined by the three data types are in fact quite different, and that we should be very wary of performing a consensus clustering type approach across data types, or any analysis strategy that assumes that all three data types are getting at the same set of underlying clusters. 

\section{Discussion}
\label{sec:discuss}
Most existing work on multiple-view clustering has focused on the problem of \emph{estimation}: namely, on exploiting the availability of multiple data views in order to cluster the observations more accurately. In this paper, we have instead focused on the relatively unexplored problem of \emph{inference}: we have proposed a hypothesis test to determine whether clusterings based on multiple data views are independent or associated.

In Section \ref{sec:app}, we applied our test to the P100 Wellness Study \citep{price2017wellness}. We found strong evidence that clusterings based on clinical data and proteomic data persist over time, i.e. that the subgroups defined by the clinical data and the proteomic data are similar at different timepoints. This suggests that if we wish to identify participant subgroups based on (say) clinical data, then it may be worthwhile to apply a consensus clustering approach to the clinical data from multiple timepoints.  However, we found only modest evidence that clusterings based on different data types are dependent!  This suggests that we should be cautious about identifying participant subgroups by applying consensus clustering across multiple data types, as the clusterings underlying the distinct data types may be quite different.

Throughout this paper, we compared clusterings on $L = 2$ data views. We may also wish to compare clusterings across $L > 2$ views. Let $X^{(l)} \in \mathbb R^{p_l}$ for $1 \leq l \leq L$ be the random vectors corresponding to the $L$ views. Suppose $X^{(l)}$ are generated according to \eqref{eq:mixtureMod} for $1 \leq l \leq L$, where $(Z^{(1)}, \ldots, Z^{(L)})$ are unobserved multinomial random variables with probabilities given by $ P(Z^{(1)}=k_1,\ldots, Z^{(L)} = k_L)=\Pi_{k_1\ldots k_L},$ for $1 \leq k_l \leq K^{(l)}$ and $1 \leq l \leq L$, where the sum of $\Pi$ over all indices is 1 and $\Pi_{k_1 \ldots k_L} \geq 0$. Results analogous to Propositions~\ref{prop:reparam} and \ref{prop:marginals} hold in this setting. Thus, we can estimate the parameters in the extended model much as we did in Section \ref{sec:estimparam}, replacing the Sinkhorn-Knopp algorithm for matrix balancing with a tensor balancing algorithm (see e.g. \citealt{sugiyama2017tensor}). To test the null hypothesis that $Z^{(1)}, \ldots, Z^{(L)}$ are mutually independent, we can develop a pseudo likelihood ratio test much as we did in Section \ref{sec:ind}, where instead of permuting the observations in $X^{(2)}$ in Step 2(a) of Algorithm \ref{alg:permuteind}, we permute the observations in $X^{(2)}, \ldots, X^{(L)}$. Alternatively, one can simply test for pairwise independence between clusterings, instead of testing for mutual independence between clusterings on all views, as we did in Section \ref{sec:app}.

An R package titled \verb+multiviewtest+ is available online at \url{https://github.com/lucylgao/multiviewtest} and is forthcoming on CRAN. Code to reproduce the data analysis in Section \ref{sec:app}, and to reproduce the simulations in Sections \ref{sec:sim} and \ref{sec:meila} and in Appendix \ref{sec:suppsim}, are available online at \url{https://github.com/lucylgao/independent-clusterings-code}. 

\section*{Acknowledgments}

We thank Nathan Price and John Earls for responding to inquiries about the P100 data, and Will Fithian for a useful conversation. Lucy L. Gao received funding from the Natural Sciences and Engineering Research Council of Canada. Daniela Witten and Jacob Bien were supported by NIH Grant R01GM123993.  Jacob Bien was supported by NSF CAREER Award DMS-1653017. Daniela Witten was supported by NIH Grant DP5OD009145, NSF CAREER Award DMS-1252624, and Simons Investigator Award No. 560585. 
{\it Conflict of Interest}: None declared.

\begin{figure}[H] 
\includegraphics[scale=0.5]{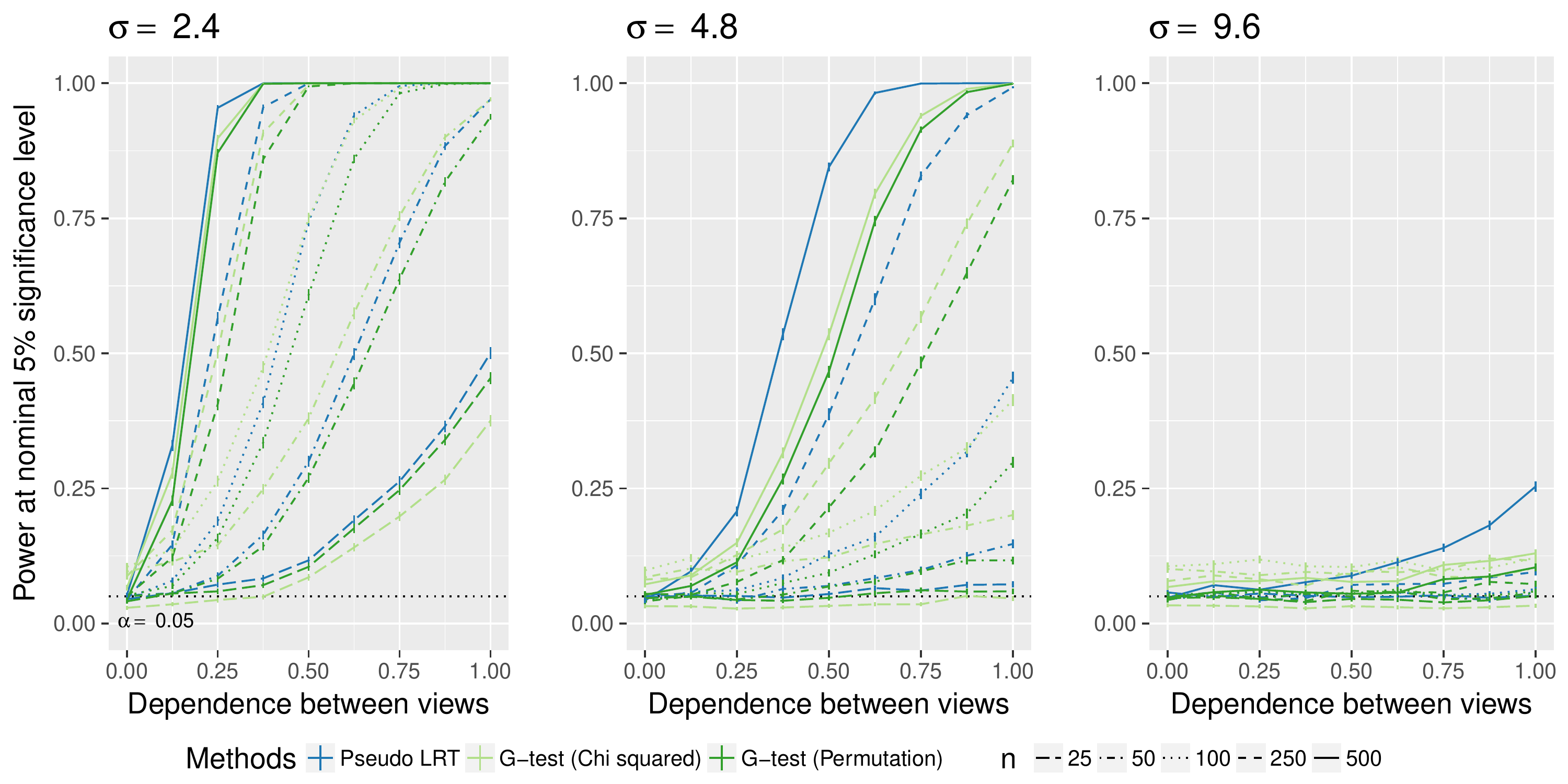}
\caption{\label{fig:meilaK6} For the simulation study described in Section \ref{sec:meila}, power of the pseudo likelihood ratio test and the $G$-test of independence for $p = 10$, $K = 6$ and  $\sigma \in \{2.4, 4.8, 9.6\}$,  with $\delta$, defined in \eqref{eq:simPi}, on the $x$-axis and power on the $y$-axis.}
\end{figure} 

 \begin{table}[H]
\centering
\begin{tabular}{ccrrrr}
View 1                     & View 2                     & \multicolumn{1}{c}{$n$} & \multicolumn{1}{c}{$p_1$} & \multicolumn{1}{c}{$p_2$} & \multicolumn{1}{c}{p-value} \\ \hline
Clinical at Timepoint 1    & Clinical at Timepoint 3    & 83                    & 204                    & 198                    &  $< 0.0001$          \\
Proteomic at Timepoint 1     & Proteomic at Timepoint 3     & 66                    & 249                    & 257                    & $< 0.0001$           \\
Metabolomic at Timepoint 1 & Metabolomic at Timepoint 3 & 88                    & 641                    & 640                    & $< 0.0001$    \\ 
\hline
Clinical at Timepoint 1 & Proteomic at Timepoint 1     & 70                    & 204                    & 249                    & 0.236           \\
Clinical at Timepoint 2 & Proteomic at Timepoint 2     & 60                    & 205                    & 254                    & 0.091         \\
Clinical at Timepoint 3 & Proteomic at Timepoint 3     & 66                    & 198                    & 257                    & 0.950          \\
Clinical at Timepoint 1 & Metabolomic at Timepoint 1 & 98                    & 204                    & 641                    & 0.034         \\
Clinical at Timepoint 2 & Metabolomic at Timepoint 2 & 89                    & 205                    & 641                    & 0.073          \\
Clinical at Timepoint 3 & Metabolomic at Timepoint 3 & 81                    & 198                    & 640                    & 0.328      \\
Proteomic at Timepoint 1  & Metabolomic at Timepoint 1 & 72                    & 249                    & 641                    & 0.402  \\
Proteomic at Timepoint 2  & Metabolomic at Timepoint 2 & 67                    & 254                    & 641                     &  0.004     \\
Proteomic at Timepoint 3  & Metabolomic at Timepoint 3 & 73                    & 257                    & 640                    &  0.020                 
\end{tabular}
\caption{\label{tab:pval} Results from the test of $H_0: C = 1_{K^{(1)}} 1_{K^{(2)}}^T$ developed in Section \ref{sec:plrt} applied to clinical, proteomic, and metabolomic data at the first and third timepoints, and applied to pairs of data views defined by different data types. Sample sizes $n$, dimensions in each view $p_1$ and $p_2$, and p-values obtained using the permutation approximation from Section \ref{sec:approxnull} are reported. }
\end{table}

\bibliography{refs}

\renewcommand\thefigure{S\arabic{figure}}
\renewcommand\thealgorithm{S\arabic{algorithm}}
\renewcommand\thefigure{S\arabic{figure}}
\setcounter{figure}{0} 
\setcounter{algorithm}{0}
\numberwithin{equation}{section}

\newpage

\appendix

\begin{center}
{\Large {\it Are Clusterings of Multiple Data Views Independent?} \\  {\bf Supplementary Materials}}
\end{center}

\section{Proofs} 
\subsection{Proof of Proposition \ref{prop:reparam}}
\label{sec:proofreparam}
 Suppose $\pi^{(1)} \in \Delta^{K^{(1)}}_+$ and $\pi^{(2)} \in \Delta^{K^{(2)}}_+$, where $\Delta^K_+ \equiv \{ s \in \mathbb R^{K} : s_k > 0, \sum \limits_{k=1}^K s_k = 1 \}$. Let $\mathcal{C}_{\pi^{(1)}, \pi^{(2)}} = \{ C \in \mathbb R^{K^{(1)} \times K^{(2)}} : C_{kk'} \geq 0, ~ C \pi^{(2)} = 1_{K^{(1)}}, ~ C^T \pi^{(1)} = 1_{K^{(2)}}\}$. First, we show that $$ \left \{ \Pi \in \Delta^{K^{(1)} \times K^{(2)}} : \Pi 1_{K^{(2)}} = \pi^{(1)}, ~\Pi^T 1_{K^{(1)}} = \pi^{(2)} \right \} \subseteq \left \{ \mathrm{diag}(\pi^{(1)}) C \mathrm{diag}(\pi^{(2)}) : C \in  \mathcal{C}_{\pi^{(1)}, \pi^{(2)}} \right \}.$$
Suppose $\Pi \in \Delta^{K^{(1)} \times K^{(2)}} = \{ S \in \mathbb R^{K^{(1)} \times K^{(2)}} : S_{kk'} \geq 0, \sum \limits_{k=1}^{K^{(1)}} \sum \limits_{k'=1}^{K^{(2)}} S_{kk'} = 1 \}$ such that $\Pi \cdot 1_{K^{(2)}} = \pi^{(1)}$ and $\Pi^T \cdot 1_{K^{(1)}}$. Define $C \in \mathbb R^{K^{(1)} \times K^{(2)}}_+$ by $C_{kk'} = \frac{\Pi_{kk'}}{\pi^{(1)}_k \pi^{(2)}_{k'}}.$
The denominator is nonzero because $\pi^{(1)} \in \Delta^{K^{(1)}}_+$ and $\pi^{(2)} \in \Delta^{K^{(2)}}_+$. Further, $C_{kk'} \geq 0$ and $\Pi = \mathrm{diag}(\pi^{(1)}) C \mathrm{diag}(\pi^{(2)})$. Now, 
\begin{align*} 
C \pi^{(2)} = \left [ \begin{matrix} \sum \limits_{k'} C_{1k'} \pi^{(2)}_{k'} \\ \vdots \\ \sum \limits_{k'} C_{K^{(1)}k'}\pi^{(2)}_{k'} \end{matrix} \right ] = \left [ \begin{matrix} \sum \limits_{k'} \frac{\Pi_{1k'}}{\pi^{(1)}_1} \\ \vdots \\  \sum \limits_{k'} \frac{\Pi_{K^{(1)}k'}}{\pi^{(1)}_{K^{(1)}}} \end{matrix} \right ] = 1_{K^{(1)}}
\end{align*} 
since $\Pi 1_{K^{(2)}} = \pi^{(1)}$ implies that $\sum \limits_{k'} \Pi_{kk'} = \pi^{(1)}_k$. Similarly, we can show $C^T \pi^{(1)} = 1_{K^{(2)}}$. 

Next, we show that 
$$ \left \{ \Pi \in \Delta^{K^{(1)} \times K^{(2)}} : \Pi 1_{K^{(2)}} = \pi^{(1)}, ~\Pi^T 1_{K^{(1)}} = \pi^{(2)} \right \}  \supseteq \left \{ \mathrm{diag}(\pi^{(1)}) C \mathrm{diag}(\pi^{(2)}) : C \in  \mathcal{C}_{\pi^{(1)}, \pi^{(2)}} \right \}.$$
Suppose $C \in \mathbb R^{K^{(1)} \times K^{(2)}}$ such that $C_{kk'} \geq 0$, $C \pi^{(2)} = 1_{K^{(1)}}$ and $C^T \pi^{(1)} = 1_{K^{(2)}}$. Define 
$$\Pi = \mathrm{diag}(\pi^{(1)}) C \mathrm{diag}(\pi^{(2)}).$$ 
Then $\Pi_{kk'} = \pi^{(1)}_k C_{kk'} \pi^{(2)}_{k'}$. Since $\pi^{(1)}_k > 0$, $C_{kk'} \geq 0$ and $\pi^{(2)}_{k'} > 0$, it follows that $\Pi_{kk'} \geq 0$. Now, 
\begin{align*} 
\Pi 1_{K^{(2)}} &= \mathrm{diag} (\pi^{(1)}) C \mathrm{diag} (\pi^{(2)}) 1_{K^{(2)}}  \\ 
&= \mathrm{diag} (\pi^{(1)}) C  \pi^{(2)} \\ 
&= \mathrm{diag} (\pi^{(1)})  1_{K^{(1)}} = \pi^{(1)}. 
\end{align*} 
Similarly, we can show $\Pi^T 1_{K^{(1)}} = \pi^{(2)}$. Further, since $\pi^{(1)} \in \Delta_+^{K^{(1)}}$, $ \sum \limits_{kk'} \Pi_{kk'} = 1_{K^{(1)}}^T \Pi 1_{K^{(2)}} = 1_{K^{(1)}}^T \pi^{(1)} = 1,$
so  $\Pi \in \Delta^{K^{(1)} \times K^{(2)}}$. 

Hence, we have proved Proposition 1.

\subsection{Proof of Proposition \ref{prop:connectmeila}} 
\label{sec:proofconnectmeila}
Let $\sigma^2 > 0$. Suppose throughout that we fit the model \eqref{eq:mixtureMod}--\eqref{eq:defPi}, for $\phi^{(1)}(\cdot; \theta)$ the density of a $N_{p_1}(\theta, \sigma^2 I_{p_1})$ random variable and $\phi^{(2)}(\cdot; \theta)$ the density of a $N_{p_2}(\theta, \sigma^2 I_{p_2})$ random variable. This amounts to applying Gaussian mixture model-based clustering with common covariance matrix $\sigma^2 I$ to each view. 

Let $\hat \pi^{(1)}$, $\hat \pi^{(2)}$, $\hat \theta^{(1)}$, and $\hat \theta^{(2)}$ denote the maximizers of \eqref{eq:marginalloglik}, and let 
\begin{align} 
\label{eq:defr}
\hat{r}^{(1)}_i = \frac{\hat \phi^{(1)}_i}{1^T \hat \phi^{(1)}_i}, ~~~ \hat{r}^{(2)}_i = \frac{\hat \phi^{(2)}_i}{1^T \hat \phi^{(2)}_i},
\end{align} 
where $\hat \phi^{(1)}$ and $\hat \phi^{(2)}$ are defined in \eqref{eq:hphi}. 

 Since Gaussian mixture model-based clustering with common covariance matrix $\nu^2 I$ converges to k-means clustering in each view as $\nu^2 \rightarrow 0$ (see Section 14.3.7 in \citet{friedman2001elements} for details), as $\sigma^2 \rightarrow 0$, 
\begin{align} 
\label{eq:singleviewconverge}
\hat r_{ik}^{(1)} \rightarrow \mathds{1} \{\tilde{M}_i^{(1)} = k\}, ~~~
\hat r_{ik'}^{(2)} \rightarrow \mathds{1} \{\tilde{M}_i^{(2)} = k'\}, ~~~ \hat \pi^{(1)}_k \rightarrow \frac{\tilde{N}_{k.}}{n}, ~~~
\hat \pi^{(2)}_{k'} \rightarrow \frac{\tilde{N}_{.k'}}{n}, 
\end{align} 
where $\tilde{M}_i^{(1)}$ and $\tilde{M}_i^{(2)}$ are the estimated k-means cluster assignments of the $i$th observation in each view, $\tilde{N}_{k.} = \sum \limits_{k'} \tilde N_{kk'}$, and $\tilde{N}_{.k'} = \sum \limits_{k} \tilde N_{kk'}$, for 
$\tilde{N}_{kk'} = | \{i \in \{1, \ldots, n\} : \tilde{M}_i^{(1)} = k, \tilde{M}_i^{(2)} = k' \} |.$

We now rewrite \eqref{eq:Copt} as 
\begin{align*} 
\hat C &= \underset{C \in \mathcal{C}_{\hat \pi^{(1)}, \hat \pi^{(2)}}}{\text{arg min}} ~ \left [ -\sum \limits_{i=1}^n \log \left ( \sum \limits_{k=1}^{K^{(1)}} \sum \limits_{k'=1}^{K^{(2)}}  \hat \pi^{(1)}_k C_{kk'}  \hat \pi^{(2)}_{k'} \hat \phi^{(1)}_{ik} \hat \phi^{(2)}_{ik'} \right )  \right ] \nonumber\\ 
&= \underset{C \in \mathcal{C}_{\hat \pi^{(1)}, \hat \pi^{(2)}}}{\text{arg min}} ~\left [ -\sum \limits_{i=1}^n \log \left ( \sum \limits_{k=1}^{K^{(1)}} \sum \limits_{k'=1}^{K^{(2)}} \hat \pi^{(1)}_k C_{kk'} \hat \pi^{(2)}_{k'} \hat \phi^{(1)}_{ik} \hat \phi^{(2)}_{ik'} \right ) - \sum \limits_{i=1}^n \log \left (1^T \hat \phi^{(1)}_i 1^T \hat \phi^{(2)}_i \right ) \right ]\nonumber\\ 
&= \underset{C \in \mathcal{C}_{\hat \pi^{(1)}, \hat \pi^{(2)}}}{\text{arg min}} ~ \left [ -\sum \limits_{i=1}^n \log \left ( \sum \limits_{k=1}^{K^{(1)}} \sum \limits_{k'=1}^{K^{(2)}} \hat \pi^{(1)}_k C_{kk'} \hat \pi^{(2)}_{k'} \hat r^{(1)}_{ik}\hat r^{(2)}_{ik'} \right ) \right ],
\end{align*} 
where the second equality holds because the quantities $\hat \phi^{(1)}_i$ and $\hat \phi^{(2)}_i$ defined in \eqref{eq:hphi} do not depend on $C$, and the third equality follows from the definition of $\hat r^{(1)}_i$ and $\hat r^{(2)}_i$ in \eqref{eq:defr}. It follows that 
\begin{align} 
\label{eq:defCmin}
\hat{C} = \underset{C \in  \mathcal{C}_{\hat \pi^{(1)}, \hat \pi^{(2)}}}{\text{arg min}} ~ g^\sigma(C),
\end{align} 
where 
\begin{align*} 
g^\sigma(C) \equiv- \sum \limits_{i=1}^n \log \left ( \sum \limits_{k=1}^{K^{(1)}} \sum \limits_{k'=1}^{K^{(2)}} \hat \pi^{(1)}_k C_{kk'}\hat \pi^{(2)}_{k'}\hat r^{(1)}_{ik}\hat r^{(2)}_{ik'} \right ).
\end{align*} 
By \eqref{eq:singleviewconverge}, as $\sigma^2 \rightarrow 0$, $g^\sigma$ converges pointwise to $g$, where 
\begin{align*} 
g(C) &\equiv -\sum \limits_{i=1}^n \log \left ( \sum \limits_{k=1}^{K^{(1)}} \sum \limits_{k'=1}^{K^{(2)}} \frac{\tilde{N}_{k.}}{n}  C_{kk'}\frac{\tilde{N}_{.k'}}{n} \mathds{1} \{\tilde M^{(1)}_i = k \} \mathds{1} \{ \tilde M^{(2)}_i = k' \}  \right ) \\ 
&= - \sum \limits_{i=1}^n \sum \limits_{k=1}^{K^{(1)}} \sum \limits_{k'=1}^{K^{(2)}} \mathds{1} \{\tilde M^{(1)}_i = k \} \mathds{1} \{ \tilde M^{(2)}_i = k' \} \log \left (\frac{\tilde{N}_{k.}}{n}  C_{kk'}\frac{\tilde{N}_{.k'}}{n}\right ) \nonumber \\ 
&= - \sum \limits_{k=1}^{K^{(1)}} \sum \limits_{k=1}^{K^{(2)}}  \tilde{N}_{kk'} \log \left (\frac{\tilde{N}_{k.}}{n}C_{kk'} \frac{\tilde{N}_{.k'}}{n} \right ).
\end{align*} 
Applying the method of Lagrange multipliers, we find that $\tilde{C} \equiv  \underset{C \in \mathcal{C}_{\hat \pi^{(1)}, \hat \pi^{(2)}}}{\text{arg min}} ~ g(C)$ 
satisfies
\begin{align} 
\label{eq:solvelimitop} 
\tilde{C}_{kk'} = \frac{n\tilde{N}_{kk'}}{\tilde{N}_{k.}\tilde{N}_{.k'}}.
 \end{align} 
 By Exercise 7.23(c) in \citet{rockafellar1998variational}, $\{ g^\sigma(C) \}_{\{\sigma > 0\}}$ is essentially bounded. By Theorem 7.17 in \citet{rockafellar1998variational}, the epigraphical limit of $g^\sigma$ is $g$. Finally, $g^\sigma$ and $g$ are continuous and proper. Hence, by Theorem 7.33 in \citet{rockafellar1998variational}, 
\begin{align}
\label{eq:defCtilde}
 \tilde{C} = \underset{C \in \mathcal{C}_{\hat \pi^{(1)}, \hat \pi^{(2)}}}{\text{arg min}} ~ g(C) = \underset{\sigma \rightarrow 0}{\lim} ~ \underset{C \in \mathcal{C}_{\hat \pi^{(1)}, \hat \pi^{(2)}}}{\text{arg min}} ~ g^\sigma(C).
\end{align} 
By \eqref{eq:defCmin}, \eqref{eq:solvelimitop} and \eqref{eq:defCtilde}, as $\sigma^2 \rightarrow 0$ we have 
\begin{align} \hat{C}_{kk'} \rightarrow \tilde{C}_{kk'} = \frac{n\tilde{N}_{kk'}}{\tilde{N}_{k.}\tilde{N}_{.k'}}.\label{eq:Cconverge}
\end{align}
By \eqref{eq:singleviewconverge} and \eqref{eq:Cconverge}, and the definition of $\hat \Pi$ in Algorithm \ref{alg:estparam}, 
\begin{align}
\label{eq:convergePi} 
\hat \Pi_{kk'} \rightarrow \frac{\tilde{N}_{kk'}}{n}.
\end{align} 
Applying \eqref{eq:singleviewconverge} and \eqref{eq:convergePi} to \eqref{eq:softstat} yields the result. 

\section{Exponentiated gradient descent for solving \eqref{eq:Copt}}
\label{sec:eg}
After a transformation of the optimization problem, \eqref{eq:Copt} can be efficiently solved using
exponentiated gradient descent \citep{kivinen1997exponentiated}, a first-order method specially
designed for optimization over the simplex. This is a form of mirror
descent, with provable convergence results \citep{beck2003mirror}. While there is no analytic solution for the update performed at each iteration of the exponentiated gradient descent, each update can be performed by applying the Sinkhorn-Knopp algorithm, a matrix balancing algorithm with provable convergence results and linear convergence rates \citep{franklin1989scaling}. 

\medskip

Define
$$ \ell(\theta^{(1)}, \theta^{(2)}, \Pi) = \sum \limits_{i=1}^n \log f(X^{(1)}_i, X^{(2)}_i; \theta^{(1)}, \theta^{(2)}, \Pi)$$ 
where $f(\cdot, \cdot; \theta^{(1)}, \theta^{(2)}, \Pi)$ is defined in \eqref{eq:jointdensity}. Consider the following optimization problem: 
\begin{align} 
\label{eq:Piopt} 
\begin{aligned}
& \underset{\Pi}{\text{minimize}}
& &  - \ell(\hat \theta^{(1)},\hat \theta^{(2)},\Pi)\\
& \text{subject to}
& & \Pi 1_{K^{(2)}} = \hat \pi^{(1)}  \\
& & & \Pi^T 1_{K^{(1)}} = \hat \pi^{(2)} \\ 
& & & \Pi_{kk'} \ge0.
\end{aligned}
\end{align} 
By Proposition \ref{prop:reparam}, we can equivalently write \eqref{eq:Piopt} as follows: 
\begin{align*} 
\begin{aligned}
& \underset{C}{\text{minimize}}
& &  - \ell(\hat \theta^{(1)},\hat \theta^{(2)},\hat \pi^{(1)}, \hat \pi^{(2)}, C)\\
& \text{subject to}
& & \mathrm{diag}(\hat \pi^{(1)}) C \mathrm{diag}(\hat \pi^{(2)}) 1_{K^{(2)}}  = \hat \pi^{(1)}  \\
& & & \mathrm{diag}(\hat \pi^{(2)}) C^T \mathrm{diag}(\hat \pi^{(1)}) 1_{K^{(1)}}  = \hat \pi^{(2)} \\ 
& & & C_{kk'} \ge0
\end{aligned}
\end{align*} 
which is equivalent to
\begin{align*} 
\begin{aligned}
& \underset{C}{\text{minimize}}
& &  - \ell(\hat \theta^{(1)},\hat \theta^{(2)},\hat \pi^{(1)}, \hat \pi^{(2)}, C)\\
& \text{subject to}
& & \mathrm{diag}(\hat \pi^{(1)}) C \hat \pi^{(2)} = \hat \pi^{(1)}  \\
& & & \mathrm{diag}(\hat \pi^{(2)}) C^T \hat \pi^{(1)}  = \hat \pi^{(2)} \\ 
& & & C_{kk'} \ge0
\end{aligned}
\end{align*} 
which is in turn equivalent to
\begin{align*} 
\begin{aligned}
& \underset{C}{\text{minimize}}
& & - l(\hat \theta^{(1)}, \hat \theta^{(2)}, \hat \pi^{(1)}, \hat \pi^{(2)}, C) \\
& \text{subject to}
& & C \hat \pi^{(2)} = 1_{K^{(1)}} \\ 
& & &  C^T \hat \pi^{(1)} = 1_{K^{(2)}} \\ 
& & & C_{kk'} \geq 0,
\end{aligned} 
\end{align*} 
which is \eqref{eq:Copt}, the optimization problem we must solve to estimate $\hat{C}$. Hence, to find $\hat{C}$, we can solve \eqref{eq:Piopt}; let $\hat \Pi$ be the minimizer of \eqref{eq:Piopt}. Then, $\hat{C}$ can be found by
\begin{align} 
\label{eq:Chat}
\hat{C}_{kk'} = \frac{\hat \Pi_{kk'}}{\hat \pi^{(1)}_k \hat \pi^{(2)}_{k'}}.
\end{align} 
The motivation for this transformation of \eqref{eq:Copt} is that the
transformed problem \eqref{eq:Piopt} can be efficiently solved using an algorithm
described in \citet{cuturi2013sinkhorn}. 

We will describe the exponentiated gradient algorithm to solve the
general problem 
\begin{align} 
\label{eq:genexpgrad}
\begin{aligned}
& \underset{\Pi}{\text{minimize}}
& &  g(\Pi) \\
& \text{subject to}
& & \sum \limits_{k'} \Pi_{kk'} = \hat \pi^{(1)}_{k}  \\
& & & \sum \limits_{k} \Pi_{kk'} = \hat \pi^{(2)}_{k'} \\ 
& & & \Pi_{kk'} \ge0
\end{aligned}.
\end{align} 
To solve \eqref{eq:genexpgrad}, we apply the update
\begin{align} 
\label{eq:Piupdateopt}
\hat \Pi^{t+1} = \begin{cases} 
\begin{aligned}
& \underset{\Pi}{\text{arg min}}
& &  g(\hat \Pi^{t})+\langle\nabla g(\hat \Pi^{t}),\Pi-\hat \Pi^{t}\rangle+\frac1{s}\sum_{kk'}\Pi_{kk'}\log(\Pi_{kk'}/\hat \Pi^{t}_{kk'}) \\
& \text{subject to}
& & \sum \limits_{k'} \Pi_{kk'} = \hat \pi^{(1)}_{k}  \\
& & & \sum \limits_{k} \Pi_{kk'} = \hat \pi^{(2)}_{k'} \\ 
& & & \Pi_{kk'} \ge0.
\end{aligned}
\end{cases}
\end{align} 
This is similar to the proximal gradient method, but instead of
$\|\Pi-\hat \Pi^{t}\|_F^2/(2s)$ we use the Bregman divergence,
$$
\frac1{s}\sum_{kk'}\Pi_{kk'}\log(\Pi_{kk'}/\hat \Pi^{t}_{kk'}).
$$
An advantage of this choice is that the positivity constraint is
automatically enforced.
For more on this, see \citet{beck2003mirror}. 
The optimality conditions for the problem \eqref{eq:Piupdateopt} are 
\begin{align} 
\label{eq:optcond}
[\nabla g(\hat \Pi^{t})]_{kk'}+[1+\log(\hat \Pi^{t+1}_{kk'}/\hat \Pi^{t}_{kk'})]/s+\lambda_k + \eta_{k'}=0,
\end{align}
for $1 \leq k \leq K^{(1)}$, $1 \leq k' \leq K^{(2)}$, where $\lambda_k$ and $\eta_{k'}$ are Lagrange multipliers for the row sum and column sum constraints, respectively. This implies that
$$
\hat \Pi^{t+1}_{kk'}=\hat \Pi^{t}_{kk'}\exp\{-s\lambda_k - s \eta_{k'} -s[\nabla g(\hat \Pi^{t})]_{kk'}-1\}.
$$
The gradient is given by
\begin{align} 
\label{eq:grad}
\nabla g(\Pi) = -\sum_{i=1}^n\frac{\hat \phi^{(2)}_i[\hat \phi^{(1)}_i]^T}{[\hat \phi^{(1)}_i]^T\Pi\hat \phi^{(2)}_i}.
\end{align} 
In the special case of \eqref{eq:Piopt}, writing $G_{kk'}=[\nabla g(\hat \Pi^{t})]_{kk'},$ 
the update
\begin{align*} 
\hat \Pi^{t+1}_{kk'}= \hat \Pi^{t}_{kk'}\exp\{s G_{kk'} - 1\} \exp \{-s \lambda_k \} \exp \{-s\eta_{k'} \}
\end{align*} 
can be written as
\begin{align*} 
\hat \Pi^{t+1} = \mathrm{diag}(v) M \mathrm{diag}(u), 
\end{align*} 
where $u_k = \exp \{-s \lambda_k \}$, $v_{k'} = \exp \{-s \eta_{k'} \}$, and $M_{kk'} = \hat \Pi^{t}_{kk'}\exp\{sG_{kk'} - 1\}$. 

\medskip 

Since $\hat \Pi^{t+1}$ must satisfy the row and column sum constraints, $u$ and
$v$ must be chosen accordingly. As in \citet{cuturi2013sinkhorn}, we
can apply the Sinkhorn Theorem and Sinkhorn-Knopp algorithm to find
$u$ and $v$. By the Sinkhorn Theorem, $u$ and $v$ are unique modulo
scalar multiplication of $u$ with a positive number and scalar
division of $v$ by that same positive number. The Sinkhorn-Knopp algorithm
(alternatively rescaling the columns so that the rows sum to
$\hat \pi^{(1)}$ then rescaling the rows so that the columns sum to
$\hat \pi^{(2)}$) can be applied to $M_{kk'}$ to find $u$ and $v$. Hence,
to perform the update, we simply multiply $\hat \Pi^{t}_{kk'}$ by
$\exp\{sG_{kk'}- 1 \}$ and then apply the Sinkhorn-Knopp algorithm to
the updated matrix so that the row and column sum constraints are
satisfied.  Algorithm \ref{alg:SK-for-Pi} provides the details. Using Proposition \eqref{prop:reparam}, we can then use our bijection between $\Pi$ to $C$  \eqref{eq:Chat} to obtain Algorithm \ref{alg:estparam} from Algorithm \ref{alg:SK-for-Pi}. 
\begin{algorithm}[t]
\caption{Exponentiated Gradient Descent for Solving  \eqref{eq:Piopt}}
    \begin{enumerate}
    \item Choose a fixed step size $s$ and compute $\hat \phi^{(1)}$ and
      $\hat \phi^{(2)}$ according to \eqref{eq:hphi}.
\item For $t=1,2,\ldots$, until
    convergence,
    \begin{enumerate} 
    \item Define 
    $$M_{kk'} = \hat \Pi^{t}_{kk'}\exp\{sG_{kk'} - 1\}$$
where 
$$
G_{kk'}=\sum_{i=1}^n\frac{\hat \phi^{(1)}_{ik}\hat \phi^{(2)}_{ik'}}{[\hat \phi^{(1)}_i]^T\hat \Pi^{t}\hat \phi^{(2)}_i}.
$$
\item (Sinkhorn-Knopp algorithm)  Let $u^0 = 1_{K^{(2)}}$ and $v^0 = 1_{K^{(1)}}$. For $t' = 1, 2, \ldots$, until convergence,
\begin{itemize} 
\item $u^{t'} = \frac{\hat \pi^{(2)}}{(\hat \Pi^{t+1})^T v^{t' - 1}}$, where the fraction denotes element-wise vector division
\item $v^{t'} = \frac{\hat \pi^{(1)}}{\hat \Pi^{t+1} u^{t'}}$,   where the fraction denotes element-wise vector division
\end{itemize} 
\item Let $u$ and $v$ denote the vectors to which $u^{t'}$ and $v^{t'}$ converge. Update
$$\hat \Pi^{t+1}_{kk'}=  u_k M_{kk'} v_{k'}. $$ 
    \end{enumerate}
\end{enumerate}
\label{alg:SK-for-Pi}
\end{algorithm}

\section{Simulations}
\label{sec:suppsim}
\subsection{Mean matrices for simulations in Section \ref{sec:sim}}
\label{sec:meanmat}
In the simulations described in Section \ref{sec:sim}, data are generated from \eqref{eq:mixtureMod}--\eqref{eq:defPi} with 
\begin{align} \label{eq:simPisupp}
\Pi = \frac{1-\delta}{K^2} 1_K 1_K^T + \frac{\delta}{K} I_K
\end{align}  for $K = 6$. In the $l$th data view, the observations are drawn from a multivariate Gaussian mixture model, for which the $k$th component in the mixture (corresponding to the $k$th cluster) is a $N_p(\mu^{(l)}_k, \sigma^2 I_p)$ distribution, with $p = 10$. The $p \times K$ mean matrices for the two data views are of the form
\begin{align*} 
\mu^{(1)} &= \left [ \begin{matrix} 2\cdot1_{5} & 0_{5} & 2 \cdot1_{5} & -2 \cdot 1_{5} & 0_{5} & -2 \cdot 1_{5}   \\ 
0_{5} & 2\cdot 1_{5} & -2\cdot 1_{5} & 0_{5} & -2 \cdot 1_{5} & 2 \cdot 1_{5} \end{matrix} \right ],  \\ 
\mu^{(2)} &= \left [\begin{matrix} -2 \cdot1_{6} & 0_{6}  & -2 \cdot1_{6}  & 2 \cdot 1_{6} & 0_4  & 2 \cdot 1_{4} \\ 0_{4} & -2 \cdot1_{4} & 2 \cdot1_{4} & 0_4 & 2 \cdot 1_{6} & -2 \cdot 1_{6} \end{matrix} \right ].
\end{align*} 

\subsection{Supplementary simulations to Section \ref{sec:sim}}
\label{sec:suppsim1}
\subsubsection{Selection of the number of clusters}
\label{sec:chooseK}

Recall from Section \ref{sec:sim} that using too few clusters in the pseudo likelihood ratio test sometimes yields better power than using the correct number of clusters. In this section, we demonstrate that using too few clusters can either increase or decrease the power, depending on the situation.

\begin{figure}[H] 
\begin{center}  
\includegraphics[scale=0.5]{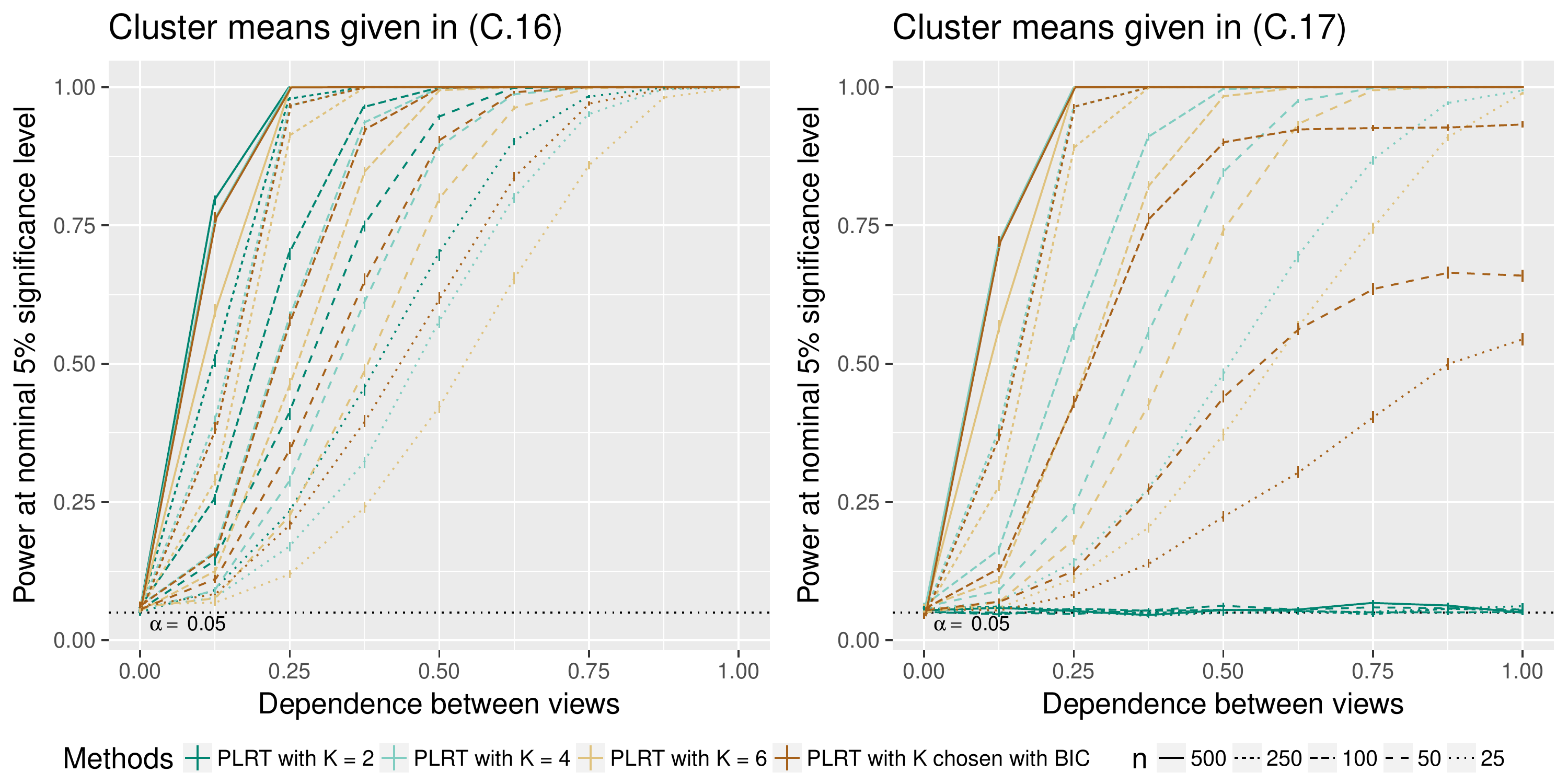}
\caption{\label{fig:compare-K} Power of the pseudo likelihood ratio test of $H_0: C = 1_{K_1} 1_{K_2}^T$ with $p = 2$, $K = 4$, and $\sigma = 0.4$ in the two simulation settings described in Appendix \ref{sec:chooseK}. The $x$-axis displays  $\delta$, defined in \eqref{eq:simPisupp}, and the $y$-axis displays the power.}
\end{center}
\end{figure}

We generate data from \eqref{eq:mixtureMod}--\eqref{eq:defPi} with $\Pi = \frac{1-\delta}{K^2} 1_K 1_K^T + \frac{\delta}{K} I_K$ for $K = 6$. In the $l$th data view, the observations are drawn from a bivariate Gaussian mixture model, for which the $k$th component in the mixture (corresponding to the $k$th cluster) is a $N_2(\mu^{(l)}_k,  0.4^2 I_p)$ distribution. We simulate 2000 datasets for a range of values of $n$, and for two choices of $\mu^{(l)}$: 
\begin{list}{}{}
\item{\emph{Choice 1:}} 
\begin{align} 
\label{eq:case1} 
\mu^{(1)} &= \left [ \begin{matrix} 2 & 2 & -2 & -2 \\ -2 & -1 & 1 & 2 \end{matrix} \right ], \quad \mu^{(2)} = \left [ \begin{matrix} -2 & -2 & 2 & 2 \\ -2 & -1 & 1 & 2 \end{matrix} \right ],
\end{align} 
\item{\emph{Choice 2:}}  \begin{align} 
\label{eq:case2}
\mu^{(1)} &= \left [ \begin{matrix} 2 & 2 & -2 & -2 \\ -2 & -1 & 1 & 2 \end{matrix} \right ], \quad \mu^{(2)} = \left [ \begin{matrix} 2 & -2 & -2 & 2 \\ 2 & -2 & -1 & 1 \end{matrix} \right ].
\end{align} 
\end{list}
We evaluate the power of the pseudo likelihood ratio test of $H_0: C = 1_{K_1} 1_{K_2}^T$ at nominal significance 
level $\alpha = 0.05$, when the number of clusters is incorrectly and correctly specified. Results are displayed in Figure \ref{fig:compare-K}. 

\begin{figure}[H]
\begin{center}  
(i)\\
\includegraphics[scale=0.5]{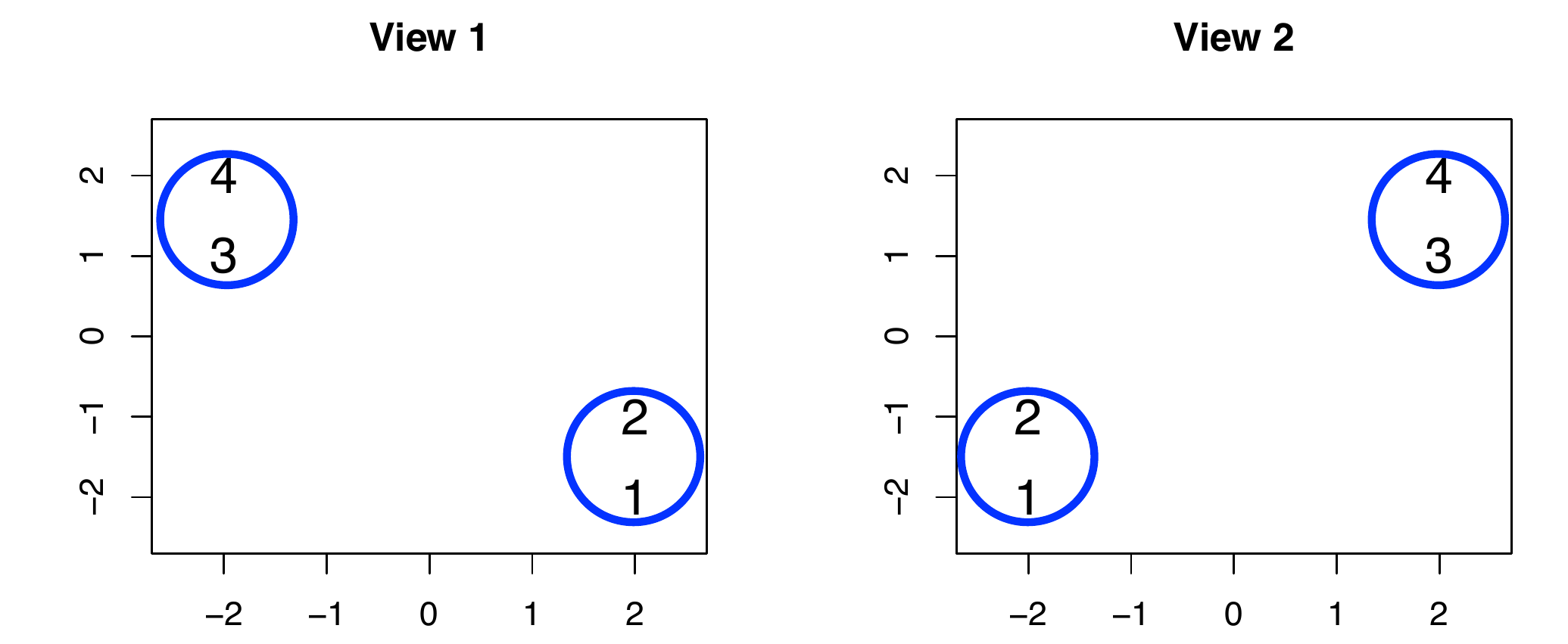} \\ 
(ii) \\ 
 \includegraphics[scale=0.5]{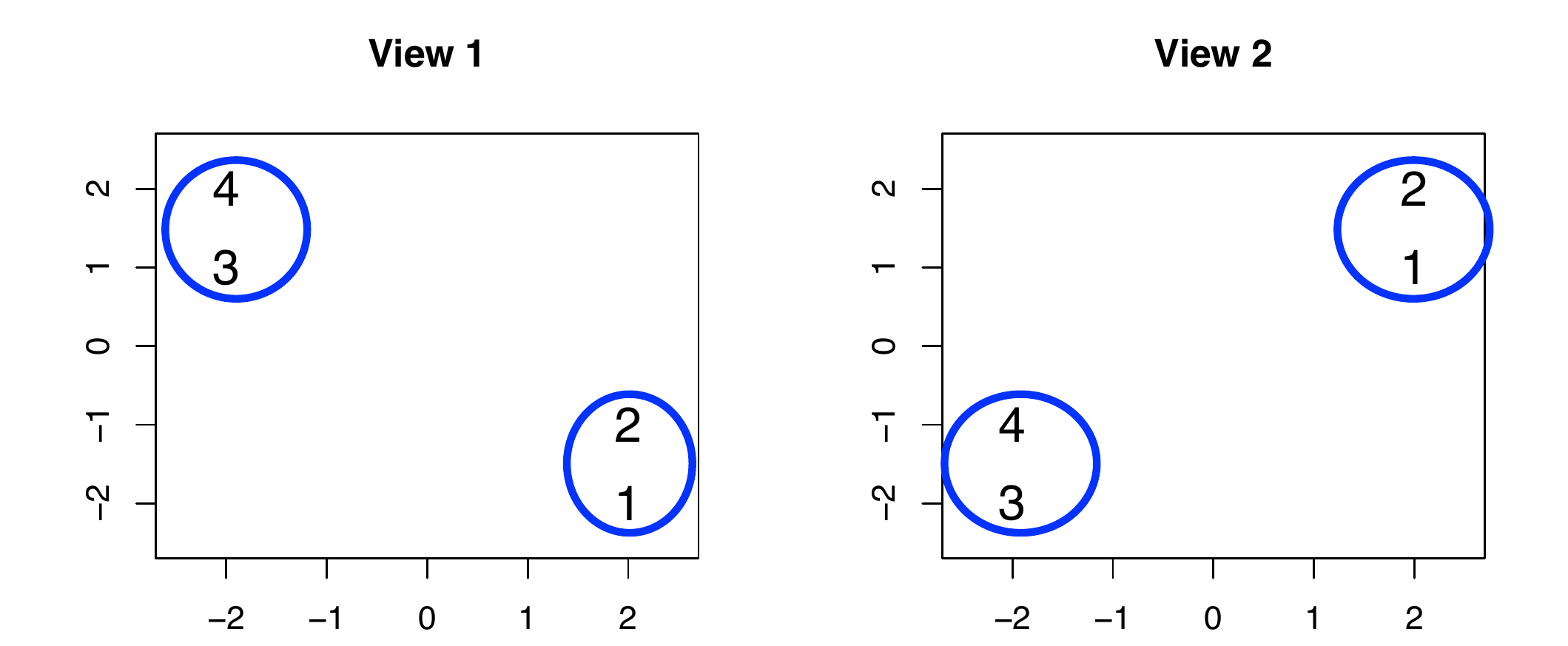}
\caption{\label{fig:meta} (i) For the simulation study described in Appendix \ref{sec:chooseK}, the cluster means and ``meta-clusters" under \eqref{eq:case1}, where the $k$th cluster mean on each view is indicated by the number $k$, and the two ``meta-clusters" on each view are circled in blue. (ii) For the simulation study described in Appendix \ref{sec:chooseK}, the cluster means and ``meta-clusters" under \eqref{eq:case2}, where the $k$th cluster mean on each view is indicated by the number $k$, and the two ``meta-clusters" on each view are circled in blue.}
\end{center}
\end{figure}

Observe from the left panel of Figure \ref{fig:compare-K} that when $\mu^{(l)}$ is given in \eqref{eq:case1}, using too few clusters $(K = 2)$ yields higher power than the correct number of clusters $(K = 4$). This is because under \eqref{eq:case1}, the clusterings on each view contain two natural ``meta-clusters", formed by combining clusters whose means are close; see Figure \ref{fig:meta}(i) for an illustration. Because the clusters on each view are not well-separated, it is easier to instead cluster the data into the two ``meta-clusters", which are highly in agreement when the four clusters are in agreement.  For example, when $\Pi = I_4/4$, the $\Pi$ matrix corresponding to the ``meta-clustering" is given by $I_2/2$. Thus, testing for independence assuming just two clusters yields better power than correctly assuming four clusters.

By contrast, observe from the right panel of Figure \ref{fig:compare-K} that when $\mu^{(l)}$ is given in \eqref{eq:case2}, using too few clusters $(K = 2)$ yields much lower power than the correct number of clusters ($K = 4)$. Again, under \eqref{eq:case2}, the clusterings on each view contain two natural ``meta-clusters" (see Figure \ref{fig:meta}(ii)), and it is easier to cluster the data into the two ``meta-clusters". However, under \eqref{eq:case2}, even when the four clusters are highly in agreement, the meta-clusters are not highly in agreement. For example, when $\Pi = I_4/4$,  the $\Pi$ matrix corresponding to the ``meta-clustering" is given by $1_2 1_2^T/4$. Thus, testing for independence assuming just two clusters yields worse power than correctly assuming four clusters. 

We also observe in both panels of Figure \ref{fig:compare-K} that using too many  clusters yields slightly lower power than using the correct number of clusters under both \eqref{eq:case1} and \eqref{eq:case2}; this result is similar to the results from the simulation setting described in Section 4 and Appendix \ref{sec:morepower1}. Furthermore, in both panels of Figure \ref{fig:compare-K}, corresponding to the two choices of $\mu^{(l)}$, choosing the clusters using BIC on each view yields slightly lower power than using the correct number of clusters in small and moderate sample sizes, and performs as well as using the correct number of clusters when the sample size is large.  

\subsubsection{Additional values of $K$ and $p$}
\label{sec:morepower1}
We simulate two data views with $p_1 = p_2 = p=10$,  and $K^{(1)} = K^{(2)} = K = 3$. In the $l$th data view, the observations are drawn from a  multivariate Gaussian mixture model, for which the $k$th component in the mixture (corresponding to the $k$th cluster) is a $N_p(\mu^{(l)}_k, \sigma^2 I_p)$  distribution.  

The means of the components  in the mixture model, written as a $p \times K$ matrix, are 
$$\mu^{(1)} = \left [ \begin{matrix} 2\cdot1_{5} & 0_{5} & 2 \cdot1_{5} \\ 
 0_{5} & 2\cdot 1_{5} & -2\cdot 1_{5} \end{matrix} \right ], ~~ \mu^{(2)} = \left [\begin{matrix} -2 \cdot1_{6} & 0_{6}  & -2 \cdot1_{6}  \\ 0_{4} & -2 \cdot1_{4} & 2 \cdot1_{4}  \end{matrix} \right ].$$ 

Additionally, we simulate two data views with $p_1 = p_2 = p=100$, and $K^{(1)} = K^{(2)} = K$, for $K = 3$ and $K = 6$. In the $l$th data view, the observations are drawn from a  multivariate Gaussian mixture model, for which the $k$th component in the mixture (corresponding to the $k$th cluster) is a $N_p(\mu^{(l)}_k, \sigma^2 I_p)$  distribution. For $K = 3$, the means of the components in the mixture model, written as a $p \times K$ matrix, are 
\begin{align*} 
\mu^{(1)} &= \left [ \begin{matrix} 2\cdot1_{50} & 0_{50} & 2 \cdot1_{50} \\ 
0_{50} & 2\cdot 1_{50} & -2\cdot 1_{50} \end{matrix} \right ],  \\ 
\mu^{(2)} &= \left [\begin{matrix} -2 \cdot1_{60} & 0_{60}  & -2 \cdot1_{60} \\ 0_{40} & -2 \cdot1_{40} & 2 \cdot1_{40} \end{matrix} \right ].
\end{align*} 
For $K = 6$, the means are 
\begin{align*} 
\mu^{(1)} &= \left [ \begin{matrix} 2\cdot1_{50} & 0_{50} & 2 \cdot1_{50} & -2 \cdot 1_{50} & 0_{50} & -2 \cdot 1_{50}   \\ 
0_{50} & 2\cdot 1_{50} & -2\cdot 1_{50} & 0_{50} & -2 \cdot 1_{50} & 2 \cdot 1_{50} \end{matrix} \right ],  \\ 
\mu^{(2)} &= \left [\begin{matrix} -2 \cdot1_{60} & 0_{60}  & -2 \cdot1_{60}  & 2 \cdot 1_{60} & 0_{40}  & 2 \cdot 1_{40} \\ 0_{40} & -2 \cdot1_{40} & 2 \cdot1_{40} & 0_{40} & 2 \cdot 1_{60} & -2 \cdot 1_{60} \end{matrix} \right ].
\end{align*} 

To investigate the type I error and power of our test,  
we generate data according to \eqref{eq:mixtureMod}--\eqref{eq:defPi}, with a range of $\Pi$ defined in \eqref{eq:simPisupp}.

\begin{figure}[H]
\centering
\includegraphics[scale=0.5]{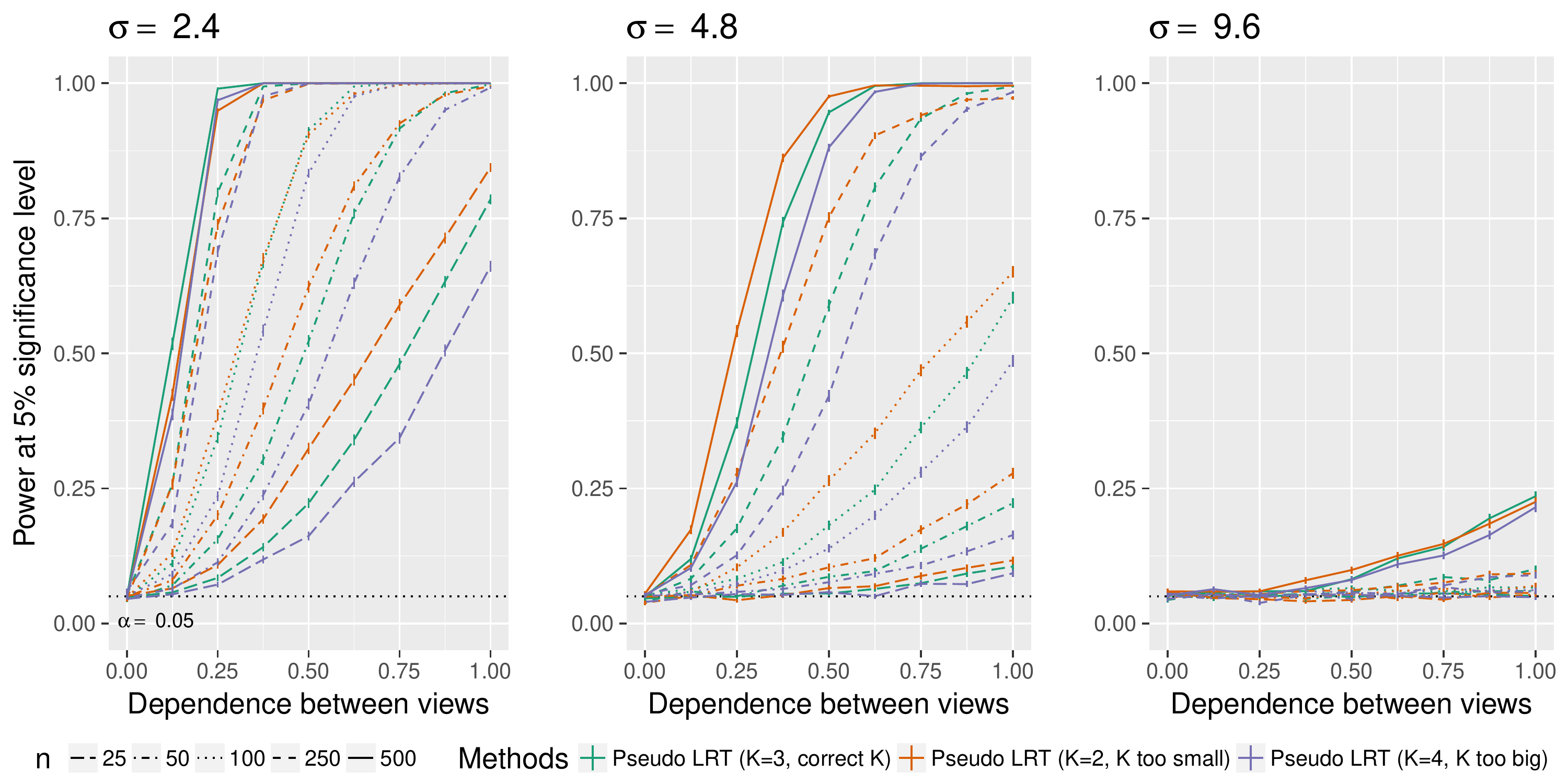}
\caption{ \label{fig:power3} Power of the pseudo likelihood ratio test of $H_0: C = 1_{K_1} 1_{K_2}^T$ with $p =10$, $K = 3$ and $\sigma \in  \{2.4, 4.8, 9.6\}$ in the simulation setting described in Appendix~\ref{sec:morepower1}. The $x$-axis displays  $\delta$, defined in \eqref{eq:simPisupp}, and the $y$-axis displays the power. }
\end{figure}

\begin{figure}[H]
\centering
\includegraphics[scale=0.5]{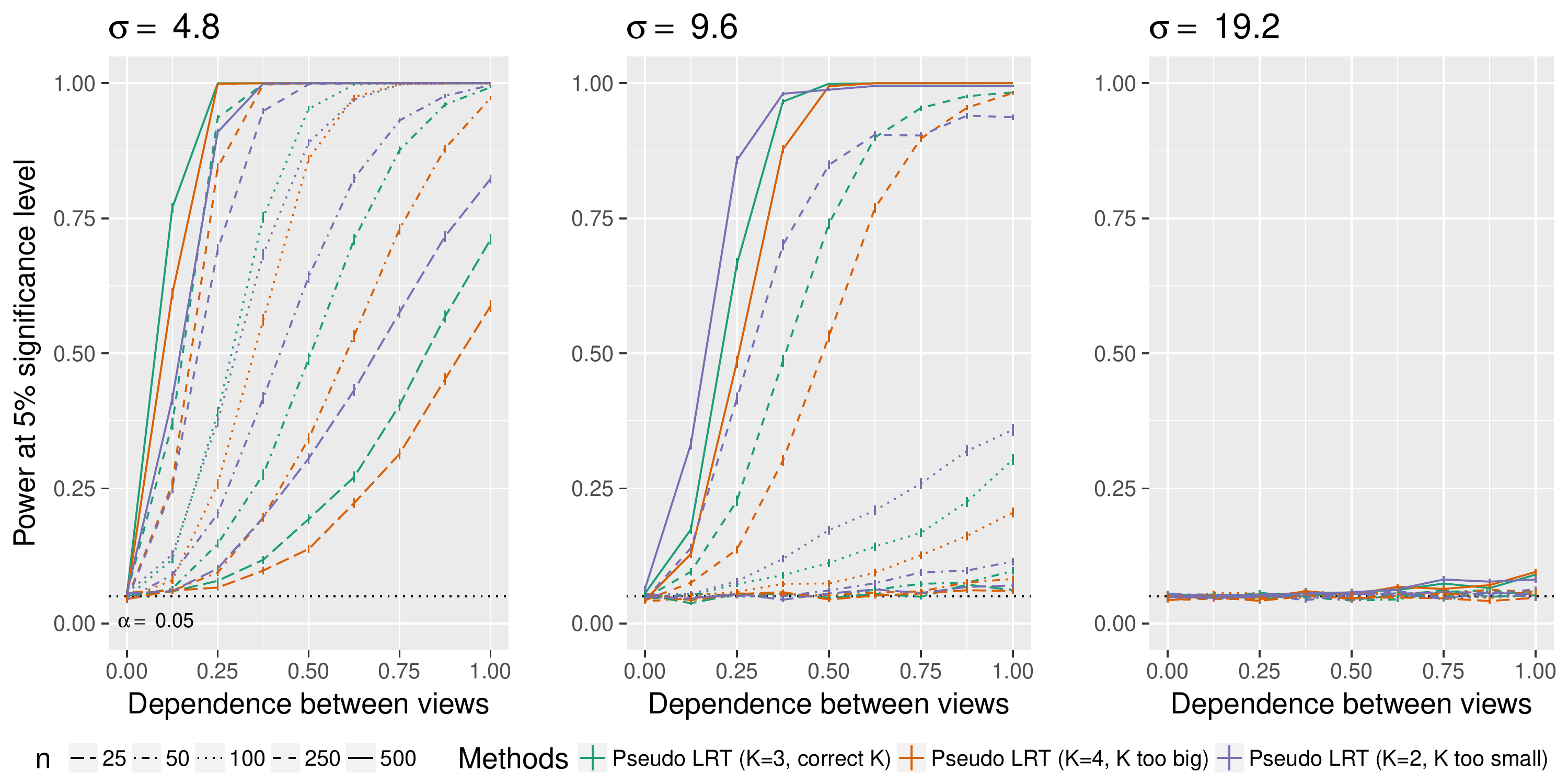}
\caption{ \label{fig:power3hd} Power of the pseudo likelihood ratio test of $H_0: C = 1_{K_1} 1_{K_2}^T$ with $p =100$, $K = 3$ and $\sigma \in  \{ 4.8, 9.6, 19.2\}$ in the simulation setting described in Appendix~\ref{sec:morepower1}. The $x$-axis displays  $\delta$, defined in \eqref{eq:simPisupp},  and the $y$-axis displays the power. }
\end{figure}

\begin{figure}[H]
\centering
\includegraphics[scale=0.5]{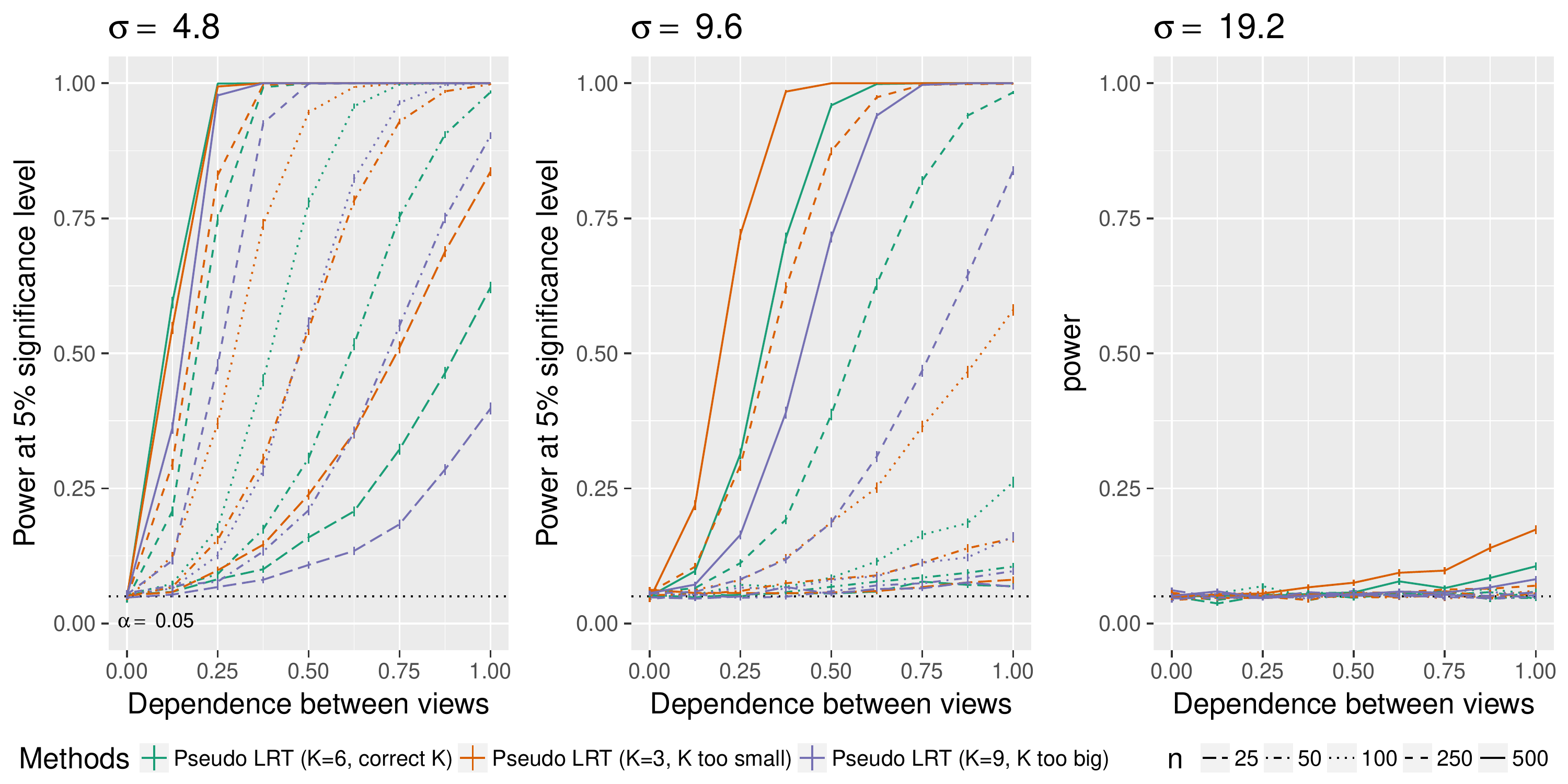}
\caption{\label{fig:power6hd} Power of the pseudo likelihood ratio test of $H_0: C = 1_{K_1} 1_{K_2}^T$ with $p = 100$, $K = 6$ and $\sigma \in  \{4.8, 9.6, 19.2\}$ in the simulation setting described in Appendix~\ref{sec:morepower1}. The $x$-axis displays  $\delta$, defined in \eqref{eq:simPisupp}, and the $y$-axis displays the power. }
\end{figure}

We simulate 2000 datasets for a range of values of $n$ and $\sigma$, and evaluate the power of the pseudo likelihood ratio test of $H_0: C = 1_{K_1} 1_{K_2}^T$ at nominal significance 
level $\alpha = 0.05$, when the number of clusters is correctly and incorrectly specified.	Results are shown in Figures~\ref{fig:power3}, \ref{fig:power3hd}, and \ref{fig:power6hd}. 

Results are similar to the simulation study described in Section \ref{sec:sim}.

\subsection{Supplementary simulations to Section \ref{sec:meila}}
\label{sec:morepower2}
In this section, we consider the $G$-test for independence in Section~\ref{sec:meila} using $\hat M^{(1)}$ and $\hat M^{(2)}$ defined in \eqref{eq:marginalclustest}. As in Section \ref{sec:meila}, we perform a simulation study in order to compare the performances of the pseudo likelihood ratio test for testing the null hypothesis $H_0: C=1_{K^{(1)}} 1_{K^{(2)}}^T$ and the $G$-test for independence. We obtain p-values for $G^2(\hat M^{(1)}, \hat M^{(2)})$ \eqref{eq:Gstat} using the $\chi^2$ approximation from \eqref{eq:chisq}, as well as a permutation approach,  where we take $B$ permutations of the elements of $\hat M^{(2)}$, and compare the observed value of $G^2(\hat M^{(1)}, \hat M^{(2)})$ to its empirical distribution in these permutation samples. 

\subsubsection{Additional values of $p$ and $K$}
\label{sec:moreK}
\begin{figure}[H]
\includegraphics[scale=0.5]{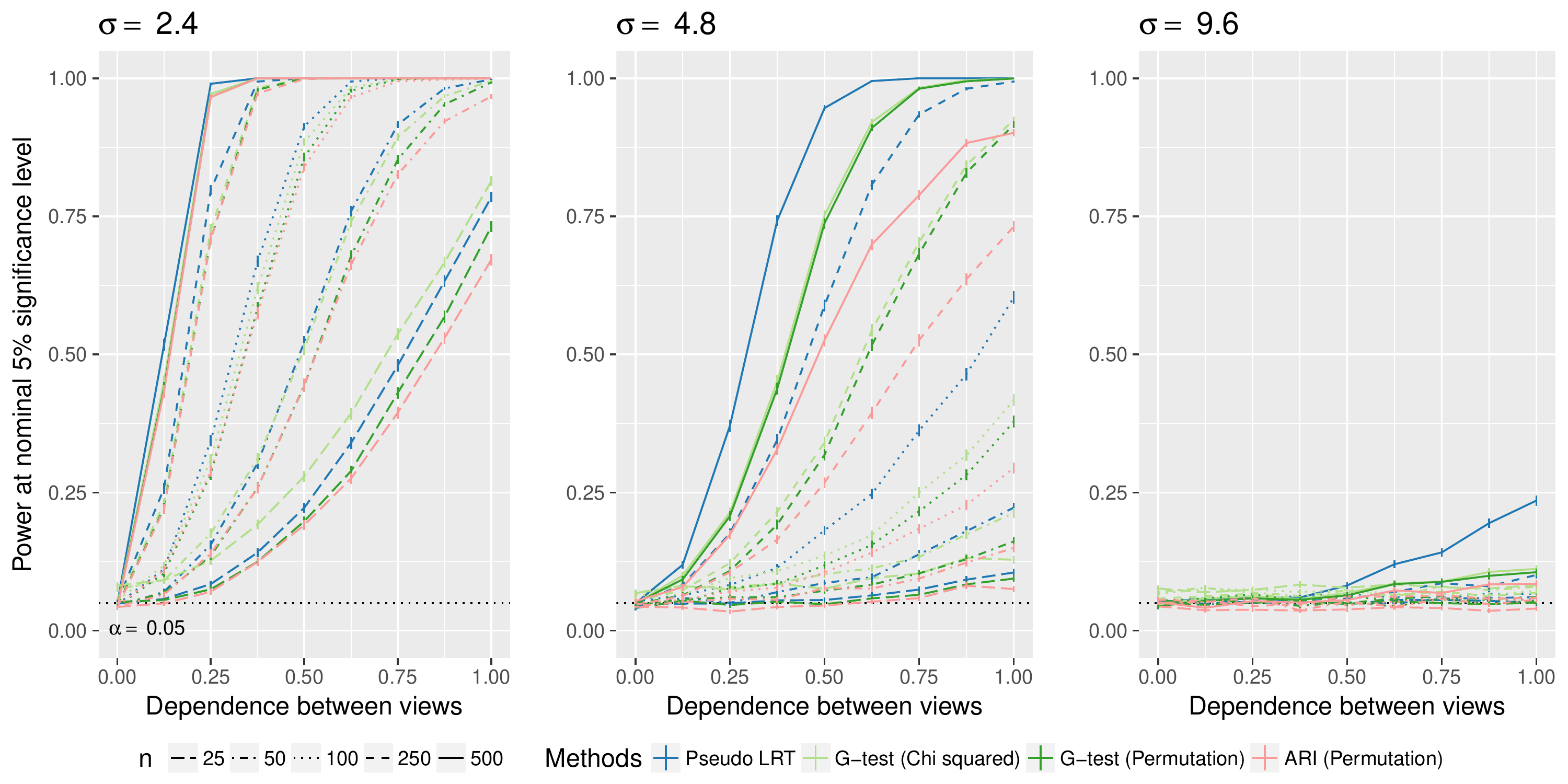}
\caption{\label{fig:meilaK3} For the simulation study described in Appendix \ref{sec:moreK}, power of the pseudo likelihood ratio test, the $G$-test for independence, and the adjusted Rand Index (ARI) for $p = 10$, $K = 3$ and  $\sigma \in \{2.4, 4.8, 9.6\}$,  with $\delta$, defined in \eqref{eq:simPisupp} on the $x$-axis and power on the $y$-axis. } 
\end{figure}

\begin{figure}[H]
\includegraphics[scale=0.5]{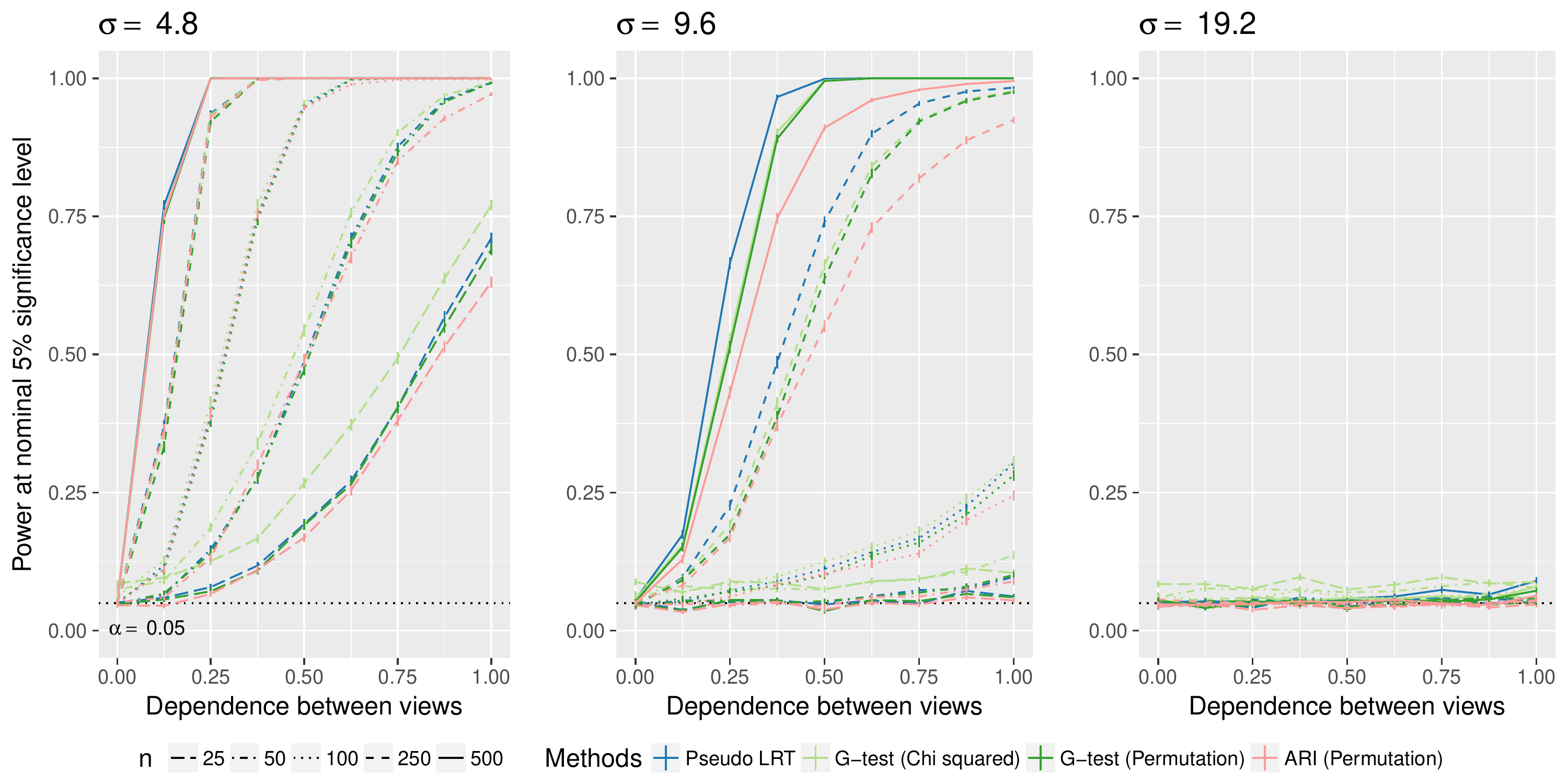}
\caption{\label{fig:meilaK3hd} For the simulation study described in Appendix \ref{sec:moreK}, power of the pseudo likelihood ratio test, the $G$-test for independence, and the adjusted Rand Index (ARI) for $p = 100$, $K = 3$ and  $\sigma \in \{4.8, 9.6, 19.2\}$,  with $\delta$, defined in \eqref{eq:simPisupp}, on the $x$-axis, and power on the $y$-axis. }
\end{figure}

\begin{figure}[H]
\includegraphics[scale=0.5]{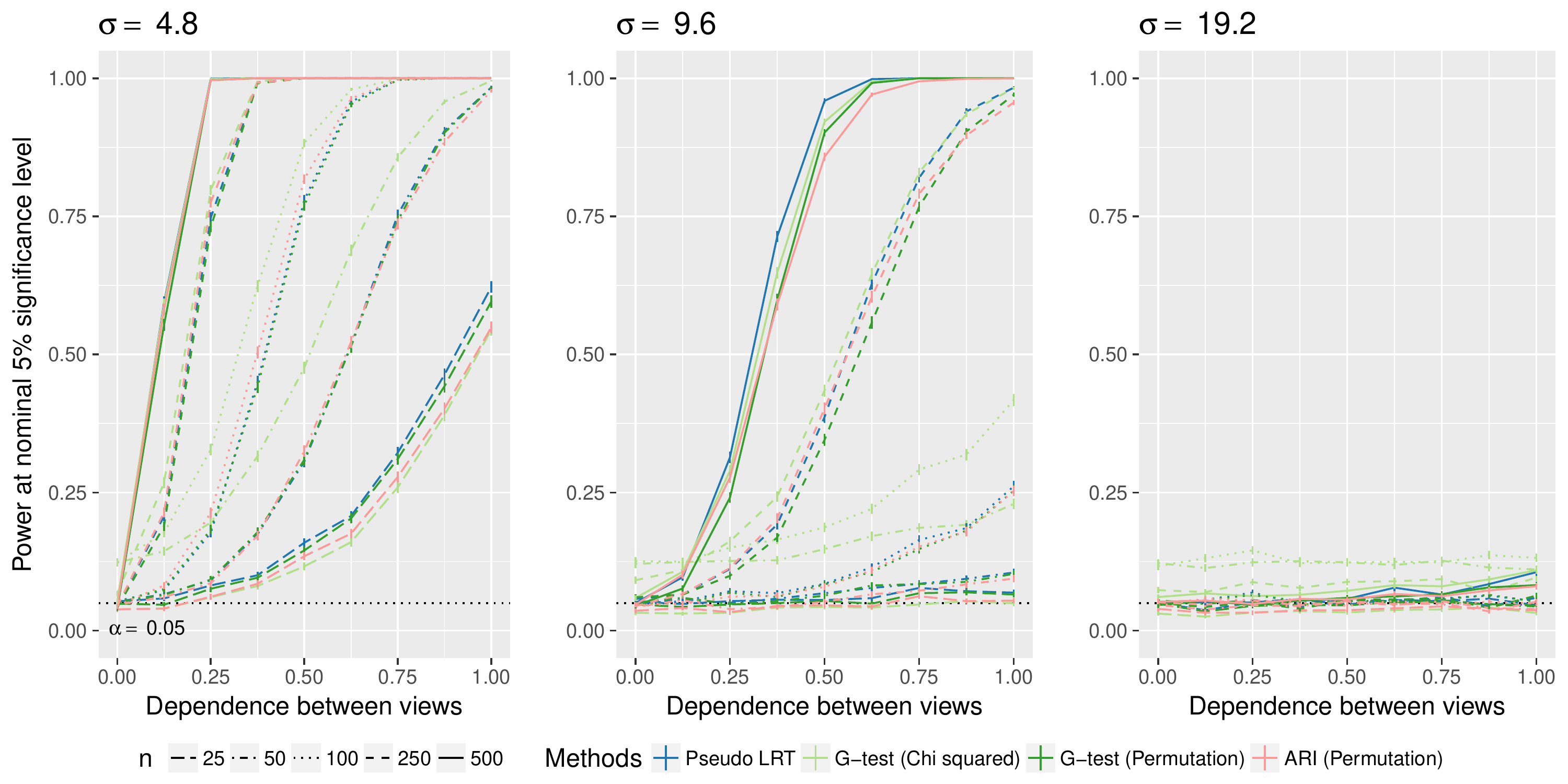}
\caption{\label{fig:meilaK6hd} For the simulation study described in Appendix \ref{sec:moreK}, power of the pseudo likelihood ratio test, the $G$-test for independence, and the adjusted Rand Index (ARI) for $p = 100$, $K = 6$ and  $\sigma \in \{4.8, 9.6, 19.2\}$,  with $\delta$, defined in \eqref{eq:simPisupp}, on the $x$-axis, and power on the $y$-axis.  }
\end{figure}

We return to the simulation set-up described in Section~\ref{sec:sim} and Appendix \ref{sec:meanmat}, with $\Sigma^{(1)} = \Sigma^{(2)} = \sigma^2 I_p$, and investigate a range of values of $p$ and $K$. In addition to the $G$-test for independence and the pseudo likelihood ratio test, we also compute the adjusted Rand Index (ARI) of \citet{hubert1985comparing} in order to compare the results of model-based clustering (implemented as in Sections \ref{sec:fitmodel} and \ref{sec:ind}) on each view; p-values for the ARI are obtained using a  permutation approach. We compare the performance of the pseudo likelihood ratio test, the $G$-test for independence, and the ARI for testing the null hypothesis $H_0: C=1_{K^{(1)}} 1_{K^{(2)}}^T$. The results are in Figures \ref{fig:meilaK3}, \ref{fig:meilaK3hd}, and \ref{fig:meilaK6hd}. Results are similar to results from the $K = 6$ and $p = 10$ setting in Figure \ref{fig:meilaK6}; we note that the ARI performs similarly to the $G$-test for independence in all cases. 

\subsubsection{Additional values of $\Sigma^{(1)}$ and $\Sigma^{(2)}$}
\label{sec:morecov}
We return to the simulation set-up described in   Section~\ref{sec:sim} with $K = 3$, $\mu^{(l)}$ given by 
\begin{align} 
\label{eq:equidistant-means}
\mu^{(1)} &= \left [ \begin{matrix} 0 & 0 & \sqrt{12} \\ 2 & -2 & 0 \end{matrix} \right ],  \quad \mu^{(2)} = \left [ \begin{matrix} -2 & 0 & 2 \\ 0 & \sqrt{12} & 0 \end{matrix} \right ],
\end{align} 
and for two choices of $\Sigma^{(l)}$: 
\begin{list}{}{}
\item{\emph{Choice 1:}} $\Sigma^{(1)} = \Sigma^{(2)} = \begin{pmatrix} 2.25 & 0.5 \\ 0.5 & 2.25 \end{pmatrix}$,
\item{\emph{Choice 2:}}  $\Sigma^{(1)} = \begin{pmatrix} 2.25 & 0.5 \\ 0.5 & 2.25 \end{pmatrix}$ and $\Sigma^{(2)} = \mathrm{diag}(2.25, 4)$.
\end{list}
To perform Step 1 of Algorithm \ref{alg:estparam}, we use the package \verb+mclust+ \citep{scrucca2016mclust} to fit Gaussian mixture models with a common dense covariance matrix (the ``EEE" covariance structure in \verb+mclust+) for $\Sigma^{(l)}$ given by choice 1 above, and to fit Gaussian mixture models with a common diagonal covariance matrix (the ``EEI" covariance structure in \verb+mclust+) for $\Sigma^{(l)}$ given by choice 2 above. We compare the performance of the pseudo likelihood ratio test of $H_0: C = 1_{K^{(1)}} 1_{K^{(2)}}^T$ at nominal significance level $\alpha$ = 0.05 to the $G$-test for independence for testing the null hypothesis $H_0: C=1_{K^{(1)}} 1_{K^{(2)}}^T$.  The results are in Figure \ref{fig:newsim}(i) and Figure \ref{fig:newsim}(ii). Results are similar to results from the $K = 6$ and $p = 10$ setting with $\sigma = 4.8$ in Figure \ref{fig:meilaK6}.

\subsubsection{Model misspecification}
\label{sec:misspec}
We consider a simulation set-up which compares the performance of the pseudo likelihood ratio test under model misspecification. We generate data from model \eqref{eq:mixtureMod} -- \eqref{eq:defPi}, with $\Pi = \frac{1-\delta}{K^2} 1_K 1_K^T + \frac{\delta}{K} I_K$ for $K = 3$. In the $l$th data view, the observations are drawn from a finite mixture model for which the $k$th component in the mixture (corresponding to the $k$th cluster) is a bivariate Student's $t$-distribution with location parameter $\mu^{(l)}_k$ and scale matrix $\Sigma = \begin{pmatrix} 2.25 & 0.5 \\ 0.5 & 2.25 \end{pmatrix}$, where the mean matrices for the two data views are of the form \eqref{eq:equidistant-means}. We fit Gaussian mixture models with a common covariance matrix (the ``EEE" covariance structure in \verb+mclust+), and again use $B = 200$ permutation samples. The results are in Figure \ref{fig:newsim}(iii). We compare the performance of the pseudo likelihood ratio test of $H_0: C = 1_{K^{(1)}} 1_{K^{(2)}}^T$ at nominal significance level $\alpha$ = 0.05 to the $G$-test for independence for testing the null hypothesis $H_0: C=1_{K^{(1)}} 1_{K^{(2)}}^T$, with p-values obtained with a permutation approach. Results remain similar to results from the $K = 6$ and $p = 10$ setting with $\sigma = 4.8$ in Figure \ref{fig:meilaK6}.

\newsavebox{\covfirst} 
\savebox{\covfirst}{$\left ( \begin{smallmatrix} 2.25 & 0.5 \\ 0.5 & 2.25 \end{smallmatrix} \right )$}

\begin{figure}[H]
\hspace{20mm} (i) \hspace{45mm} (ii) \hspace{43mm} (iii)  \\ 
\includegraphics[scale=0.5]{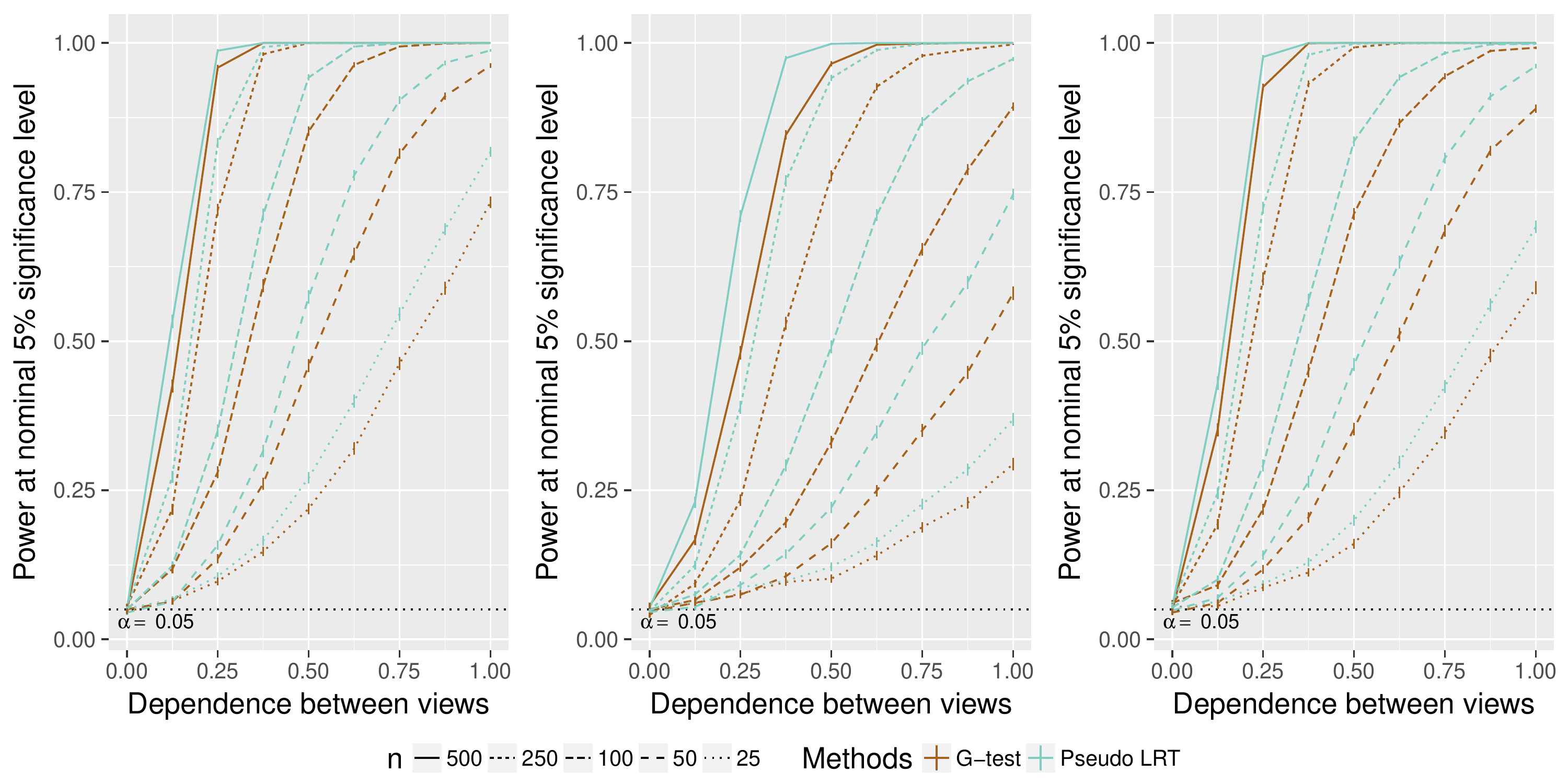}
\caption{\label{fig:newsim} The x-axis displays $\delta$, defined in \eqref{eq:simPisupp}, and the y-axis displays power. Power of the pseudo likelihood ratio test and the $G$-test for independence for 
(i) $\Sigma^{(1)} = \Sigma^{(2)} = $ \usebox{\covfirst}, and Gaussian mixture components,  
(ii)  $\Sigma^{(1)} = $ \usebox{\covfirst}, $\Sigma^{(2)} = \mathrm{diag}(2.25, 4)$, and Gaussian mixture components, (iii) bivariate Student's $t$-distributions as mixture components. Details for (i) and (ii) are in Appendix \ref{sec:morecov}, and details for (iii) are in Appendix \ref{sec:misspec}. }
\end{figure}

\end{document}